\documentclass[12pt,twoside,journal,draftclsnofoot,onecolumn]{IEEEtran}
\usepackage{amsmath,amsfonts,bm,amssymb,psfrag,ifthen,color,subfigure}
\usepackage{mathrsfs}
\usepackage[justification=centering,font=footnotesize]{caption}
\usepackage{todonotes}

\allowdisplaybreaks[4]

\usepackage[dvips]{epsfig}
\usepackage[dvips]{graphicx}
\usepackage{amsmath,amsfonts,bm,amssymb,psfrag,ifthen,color,subfigure}
\usepackage{algpseudocode}
\usepackage{algorithm}
\usepackage{algorithmicx}
\usepackage{ifthen}
\usepackage{setspace}
\usepackage{cite}
\usepackage{url}
\usepackage{longtable}
\usepackage{stfloats}  % Written by Sigitas Tolusis
\usepackage{graphics,booktabs,color,subfigure}
\usepackage{float}
\usepackage{diagbox} % ??????
\usepackage{threeparttable}
\usepackage{tabularx}

\newcommand{\tabincell}[2]{\begin{tabular}{@{}#1@{}}#2\end{tabular}}

\usepackage{listings}

\floatname{algorithm}{Algorithm}

\renewcommand{\algorithmicrequire}{ \textbf{Input:}} %Use Input in the format of Algorithm
\renewcommand{\algorithmicensure}{ \textbf{Output:}} %UseOutput in the format of Algorithm

\begin{document}
\title{ Joint  Beamforming and PD Orientation  Design for Mobile  Visible Light Communications}
\author{Shuai Ma, Jing Wang, Chun Du, Hang Li,~\IEEEmembership{Member,~IEEE}, Xiaodong Liu,  Youlong Wu,    Naofal Al-Dhahir,~\IEEEmembership{Fellow,~IEEE} and Shiyin Li
\thanks{S. Ma, J. Wang, Chun Du, and S. Li   are with the School of Information and Control   Engineering, China
University of Mining and Technology, Xuzhou, 221116,
China. (e-mail: \{mashuai001, ts19060135p31, Chun Du, lishiyin\}@cumt.edu.cn).}
%\thanks{H. Li is with the Shen zhen Research Institute of Big Data, Shenzhen 518172, Guangdong, China. (email: hangdavidli@163.com).}
%\thanks{X. Liu is with   the School of Information
%Engineering, Nanchang University, Nanchang 330031, China (e-mail: xiaodongliu@ieee.org)}
%\thanks{Y. Wu is with the School of Information Science and Technology, ShanghaiTech University, Shanghai 201210, China (email:  wuyl1@shanghaitech.edu.cn). }
%%  \thanks{M. Safari is with the School of Engineering, the University of Edinburgh, Edinburgh EH9 3JL, U.K. (e-mail: majid.safari @ed.ac.uk).}
%\thanks{N. Al-Dhahir is with the Electrical Engineering Department, University of Texas at Dallas, Dallas, TX 75080 USA (e-mail: aldhahir@utdallas.edu)}
}

\maketitle
\IEEEpeerreviewmaketitle
\begin{abstract}
In this paper, we propose   joint    beamforming and  photo-detector (PD)   orientation (BO)  optimization  schemes for mobile  visible light communication (VLC) with the orientation adjustable receiver (OAR).
Since    VLC is  sensitive to line-of-sight  propagation, we   first establish the OAR model and the human body blockage model for mobile VLC user equipment (UE).
To  guarantee the  quality of service (QoS) of mobile VLC,  we jointly optimize BO with  minimal UE the power consumption    for both  fixed  and random   UE  orientation cases.
 For the  fixed UE  orientation  case,  since the {transmit}  beamforming and  the PD   orientation are mutually coupled, the  joint    BO  optimization problem is nonconvex and intractable. To address this challenge, we propose   an alternating optimization algorithm to obtain the transmit  beamforming and  the PD   orientation.
For the random UE orientation case, we further propose a  robust  alternating BO optimization algorithm  to ensure the worst-case QoS requirement of the mobile UE.
   Finally, the performance of
joint  BO optimization design schemes are evaluated for mobile VLC through numerical
experiments.

\end{abstract}
%\newpage

\section{Introduction}

Visible light communication (VLC), which  integrates
communications and illuminations,
has recently attracted increasing attention as a promising indoor wireless
technology for the   beyond fifth-generation (B5G)    networks \cite{Wong_2017,Chettri_ITJ_2020,Agiwal_ICST_2016}.
By using  widely deployed light emitting
diodes (LEDs)  as   transmitter  antennas,  VLC  has many advantages, including being license free, huge bandwidth (380-780 THz),
high transmission
rate, and high  energy efficiency.
  Thus, VLC can
significantly alleviate the spectrum congestion of  conventional   radio-frequency (RF)
communications \cite{Abdelhady_ITC_2019}. However,
due to the inherent short wavelength characteristics of the visible light,  VLC is more susceptible to blockages than  RF communications,
resulting in  severe communication rate  drop leading to  outage {\cite{Tang_ICT_2021, Pathak_ICST_2015}}.

To   reduce the probability of being obstructed and improve  receiver   signal-to-noise ratio (SNR), there are generally two types of adjustable VLC receivers, i.e., angle diversity receiver (ADR) \cite{Nuwanpriya_JSAC_2015,chen_JLT_2018,Alsulami_ICTON_2020} and orientation adjustable receiver (OAR)      \cite{Eroglu_2021_ITMC,Abdalla_GLOBECOM_2018, Abdalla_IET_2019, Abdalla_Access_2020, Morales_Letter_2021,Alan_Sci_2018,Fan_OE_2017,Kamali_Nanophotonics_2018,Notaros_CLEO_2019,Alain_IWC_2021,Alain_JLT_2021}.
  Specifically, ADR employs    multiple photo-detectors (PDs) with different orientations  to achieve multiplexing gain\cite{Nuwanpriya_JSAC_2015}, while OAR can flexibly adjust the orientations of PDs to align with  the direction of incident VLC signals.
   Compared with ADR's reliance on multiple PDs, OAR can be implemented even with a single PD. Therefore, OAR is more flexible.
   Generally, there are two   approaches to realize OAR, i.e.,
      mechanical     control  steering  \cite{Eroglu_2021_ITMC,Abdalla_GLOBECOM_2018, Abdalla_IET_2019, Abdalla_Access_2020,Morales_Letter_2021}, and metasurfaces steering  \cite{Alan_Sci_2018,Fan_OE_2017,Kamali_Nanophotonics_2018,Notaros_CLEO_2019,Alain_IWC_2021,Alain_JLT_2021}.
On one hand, the mechanical     control  steering manipulates the orientations of PDs  by employing  mechanical  machine control.
Based on this setup,  a slow beam steering scheme \cite{Eroglu_2021_ITMC} was proposed   to maximize the  rates of the VLC system by utilizing piezoelectric actuators.
 A controlled field of
view (FOV) was exploited    to improve handover performance  \cite{Abdalla_GLOBECOM_2018},  support  device mobility \cite{Abdalla_IET_2019},  and mitigate interference\cite{Abdalla_Access_2020}.
  In \cite{Morales_Letter_2021}, a linear zero-forcing
 precoding scheme was developed for  multiuser multiple-input single-output  VLC systems, where each user could select a specific  receiving
orientation angle  from a set of possible orientations. On the other hand, the metasurfaces steering
  explores   the principle of the metasurfaces infrastructure and the physico-chemical characteristics \cite{Alan_Sci_2018,Fan_OE_2017,Kamali_Nanophotonics_2018,Notaros_CLEO_2019,Alain_IWC_2021,Alain_JLT_2021},
   to  realize focal length tuning \cite{Alan_Sci_2018} (e.g., defocus), astigmatism and shift. Such metasurface-based steering scheme   has shorter response time than the mechanical control steering scheme, and thus has attracted increasing research attention.  In \cite{Fan_OE_2017}, an autofocusing airy   beam steering scheme was designed to flexibly adjust   the focal length of visible light.
In \cite{Notaros_CLEO_2019},
liquid-crystal (LC) -based optical phased arrays were developed for visible-light beam steering.
In \cite{Alain_IWC_2021},   two types
of intelligent meta-elements to steer the incident
light beam, i.e., a meta-lens with
electrically stretchable artificial muscles and a
LC-based re-configurable intelligent surface (RIS) infrastructure with electronically
adjustable refractive index, were presented for VLC systems.
The authors in \cite{Alain_JLT_2021} further proposed  an LC-based RIS      to  enhance  the  VLC signal detection capacity and transmission range.
Given  the existing related works\cite{Alan_Sci_2018,Fan_OE_2017,Kamali_Nanophotonics_2018,Notaros_CLEO_2019,Alain_IWC_2021,Alain_JLT_2021},
 the performance of the  metasurfaces steering for VLC
 is still in the preliminary stage.

Note that   most of the exiting works {\cite{Abdalla_GLOBECOM_2018, Abdalla_IET_2019, Abdalla_Access_2020}} assume that the  PD's orientation  can be perfectly aligned with the direction of the incident light. However, for  more general practical mobile VLC scenarios, {the user equipment (UE)} may move and rotate. Thus, the  PD's orientation of the mobile UE is changing and may not always accurately be aligned with the direction of the incident light, which will lead to the deterioration of VLC performance.
    Motivated by
the above discussion,
  we focus on jointly designing the beamforming and PD orientation (BO) for  mobile VLC   with  fixed and random
UE orientation, respectively.
   Note that the proposed joint BO scheme can be applied for both   mechanical     control  steering and  metasurfaces steering  schemes.
 The main contributions of this paper are summarized as follows:

\begin{itemize}

 \item To {describe} the mobile VLC, we
 first establish  the mobile UE orientation rotation  model.
 Then, we  characterize  the channel
  blockage model of mobile  UE, which includes  specific models for OAR and  human blockage. Moreover, we analyze two types of OAR physical structures for mobile UE, i.e., mechanical control receiver and tunable LC receiver.

 \item Furthermore,  for the  fixed  UE orientation, we
jointly optimize the {transmit} beamforming vector  and the PD orientation  vector
 to minimize the total transmit power of LEDs, while satisfying the minimal rate requirement. Since the {transmit} beamforming and the
PD orientation are mutually coupled, the joint BO optimization problem is non-convex and  intractable.
 To tackle the complicated joint optimization problem, we  decompose it
into two sub-problems, i.e., the beamforming  subproblem
and  PD orientation subproblem. Then, we transform both beamforming  subproblem
and  PD orientation subproblem to convex problems, and    propose  an alternating optimization (AO) algorithm to iteratively solve the   transmit beamformer    and the PD orientation.

\item Moreover,   for the  random   UE  orientation,
we describe the random UE  orientation  model, and derive the corresponding imperfect channel state information (CSI) model.
Based on  the imperfect CSI model,
 we  further investigate  robust joint BO optimization problem to minimize the   transmit power subject to the worst-case quality of service (QoS) requirement, which is NP-hard.
 To  make the over-complicated problem tractable, we propose to
break down the robust joint BO optimization problem
into robust beamforming subproblem and PD orientation subproblem.
 For a given   UE orientation, we effectively  optimize robust beamforming     by utilizing the
semidefinite relaxation (SDR) method. With a fixed robust beamformer, the PD orientation subproblem is  {a non-linear non-smooth problem},  and we provide
an alternating optimization and projection method.

\end{itemize}

 The rest of this paper is organized as follows.  Section II describes the OAR-based system model.
    In Section III, we  present the joint BO scheme    for the fixed UE  orientation case.
   The case of   random UE orientation
 is discussed in Section IV.
        The    simulation and numerical results are  provided in Section V. Finally, the paper is concluded in    Section VI. Table I presents the means of the
key notations in  this paper.
\begin{table}[H]
	\caption{Summary of Key Notations}
	\label{tablepar}
	\centering
	\begin{tabular}{|c|l|}
		\hline
		\rule{0pt}{8pt}Notation  & Description   \\ \hline
		\rule{0pt}{7.5pt}$\theta ,\omega$ &  \tabincell{c}{Elevation  and azimuth angle of PD orientation vector} \\ \hline
		\rule{0pt}{7.5pt}${{\bf{g}}\left( {\theta ,\omega } \right)}$ &  \tabincell{c}{Channel gain function of ${\theta ,\omega }$} \\ \hline
		\rule{0pt}{7.5pt}${{\widehat{\bf n}}_{{\rm{UE}}}} $ &  \tabincell{c}{UE orientation vector in the Earth coordinate}\\ \hline
		\rule{0pt}{7.5pt}${{{\bf n}}_{{\rm{UE}}}} $ &  \tabincell{c}{UE orientation vector in the UE coordinate}\\ \hline
		\rule{0pt}{7.5pt}${{{\bf n}}_{{\rm{OAR}}}} $ &  \tabincell{c}{PD orientation vector in the UE coordinate}\\ \hline
		\rule{0pt}{7.5pt}${u_i} $ &  \tabincell{c}{Indicator parameter of blockage model}\\ \hline
		\rule{0pt}{7.5pt}${R\left( {\theta ,\omega ,{\bf{p}}} \right)} $ &  \tabincell{c}{Achievable rate of fixed UE orientation}\\ \hline
		\rule{0pt}{7.5pt}${R_{{\rm{mob}}}}\left( {{{{{\bf n}}}_{{\rm{OAR}}}}} ,{\bf{p}} \right) $ &  \tabincell{c}{Achievable rate of random UE orientation}\\ \hline
		\rule{0pt}{7.5pt}${{\bf{e}}_i}$ &  \tabincell{c}{Unit vector with the $i$-th element equal to $1$}\\ \hline
%		\rule{0pt}{7.5pt}${{\bf{1}}_N}$  &  \tabincell{c}{$N \times 1$ vector with all element equal to $1$}\\ \hline

%		\rule{0pt}{7.5pt}${{\bf{x}}_{\rm{b}}} $ &  \tabincell{c}{Intersecting point of the blocked incident light  and the human body}\\ \hline

%		\rule{0pt}{7.5pt}${\bf{R  }}$ &  \tabincell{c}{Rotated matrix of UE}\\ \hline
%		\rule{0pt}{7.5pt}${{\bf{R}}_\Delta } $ &  \tabincell{c}{Rotated matrix of random UE}\\ \hline
	\end{tabular}
\end{table}
\emph{Notations}: Boldfaced lowercase and uppercase letters represent vectors and matrices, respectively. The  transpose, Frobenius norm, Hadamard-product,  trace of a matrix and expectation are denoted as  ${\left(  \cdot  \right)^{{T}}}$,  $\left\|  \cdot  \right\|$, $\odot$,  ${\rm{Tr}}\left(  \cdot  \right)$, and $\mathbb{E}\left\{ {\cdot} \right\}$, respectively.   $\mathcal{N} \triangleq \left\{ {1,2,...,N} \right\}$.

\section{Mobile VLC System Model}

  Consider a  downlink  mobile VLC system,  where a lamp with $N$ LEDs can
 transmit information to  a  mobile UE with single PD. Let $s$ denote the transmitted signal and it follows the amplitude constraint, i.e., $\left| s \right| \le A$. Meanwhile, $\mathbb{E}\left\{ s \right\} = 0$ and $\mathbb{E}\left\{ {{s^2}} \right\} = \varepsilon $. Moreover, let {${\bf{p}} = {\left[ {\sqrt {{p_1}} ,...,\sqrt {{p_N}} } \right]^T}\in {\mathbb{R}^N}$} denote the beamforming vector, where $p_i$ is the power gain for  the $i$th LED. Thus, the transmitted signal {${\bf{x}}$} can be written as
\begin{align}\label{trans_sig}
{{\bf{x}}} = {\bf{p}}s + {I_{{\rm{DC}}}}{{\bf{1}}_N},
\end{align}
where ${I_{{\rm{DC}}}} \ge 0$ is the direct current (DC) bias at each LED, ${{\bf{1}}_N}$  denotes a
$N \times 1$ vector with all element equal to $1$.
To     ensure the non-negativity of the transmitted signal,
  the power gain $p_i$ of  the $i$th LED satisfies
\begin{align}
 \sqrt {{p_i}} A \le {I_{{\rm{DC}}}},\forall i \in {\cal N}.
 \end{align}

At the UE side, the received signal  is from both the {line-of-sight} (LOS) channel and reflection channel.
{In many scenarios, the   gains of reflection channels are much lower than those of the LOS channels \cite{Wang_JLT_2013,Fath_ITC_2013}. In this paper, we only consider cases of LOS channel dominating the transmission, i.e., the NLOS transition coefficient is less than 0.5.}
Let ${{\bf{g}}\left( {\theta ,\omega } \right)} = {\left[ {{g_1}\left( {\theta ,\omega } \right),...,{g_N}\left( {\theta ,\omega } \right)} \right]^T}$ denote  the channel gain  vector, where ${g_i}\left( {\theta ,\omega } \right)$ is the channel gain between the $i$th LED and the UE. Note that the channel gain vector ${{\bf{g}}\left( {\theta ,\omega } \right)}$ is a function of ${\theta ,\omega }$, which vary with OAR's orientation. We will specify the channel gain vector in the following subsections. By using the above definitions, the received signal $y$ at the UE is given by
 \begin{align}\label{receive_sig}
y = {{{\bf{g}}\left( {\theta ,\omega } \right)}^T}{{\bf{x}}} + z,
\end{align}
where   $z $ is the received real Gaussian noise with mean zero and variance ${{\sigma ^2}}$.

%Next, we will present the specified expression of the channel gain, followed by the achievable rate of the considered system.

%Due to the mobility of the UE, the LOS link may suffer from blockage, and the channel power gain would vary tremendously. Next, we will introduce the blockage model to further specify the channel variation.

\subsection{UE Coordination System}

\begin{figure}[htbp]
\centering
\subfigure[Horizontal.]{
\begin{minipage}[t]{0.25\linewidth}
\centering
\includegraphics[width=1.2in]{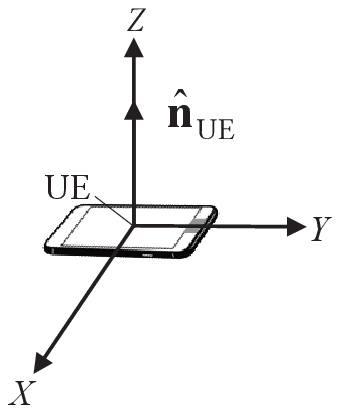}
%\caption{fig1}
\end{minipage}%
}%
\subfigure[The yaw $\alpha$.]{
\begin{minipage}[t]{0.25\linewidth}
\centering
\includegraphics[width=1.2in]{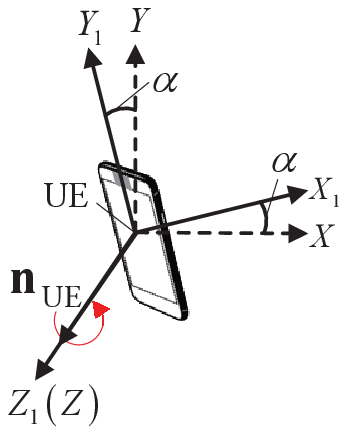}
%\caption{fig2}
\end{minipage}%
}%
\subfigure[The pitch $\beta$.]{
\begin{minipage}[t]{0.25\linewidth}
\centering
\includegraphics[width=1.2in]{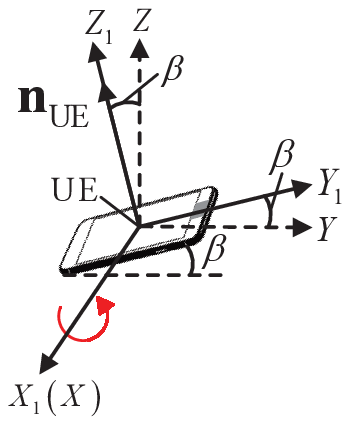}
%\caption{fig2}
\end{minipage}
}%
\subfigure[The roll $\gamma$.]{
\begin{minipage}[t]{0.25\linewidth}
\centering
\includegraphics[width=1.2in]{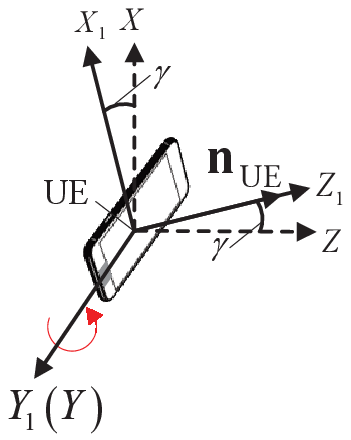}
%\caption{fig2}
\end{minipage}
}%
\centering
\caption{ The UE coordination system and its rotation.}\label{rotation_fig}
\end{figure}

To quantify the UE orientation, we first introduce the UE coordination system, as shown in Fig. 1, where
 ${{\bf{n}}_{{\rm{UE}}}}$ denotes the orientation of  the UE. When it is placed as shown in Fig. 1 (a), its direction is aligned with the Z-axis. Let ${{\widehat{\bf n}}_{{\rm{UE}}}} $ denote the orientation vector of UE in the Earth coordinate $X$-$Y$-$Z$ (in which the $X$-$Y$ plane is the horizontal plane). The UE rotation  is usually defined by the yaw, pitch, and roll. Specifically, as shown in Fig. 1 (b)-(d),
 the yaw is the positive rotation around the Z-axis with an angle of $\alpha$, pitch is  around the X-axis with an angle of $\beta$, and roll is  around the Y-axis with an angle of $\gamma$.

According to Euler's rotation theorem \cite{Zeng_2018_VTC}, the rotation   can be uniquely expressed by three elements ${{\bf{R}}_\alpha },{{\bf{R}}_\beta },{{\bf{R}}_\gamma }$, which are the yaw, pitch and roll matrices corresponding to $\alpha$, $\beta$,and $\gamma $, respectively. Specifically, the expressions for the yaw, pitch and roll matrices are given as
\begin{align}
{{\bf{R}}_\alpha } = \left[ {\begin{array}{*{20}{c}}
{\cos \alpha }&{{\rm{ - sin}}\alpha }&0\\
{{\rm{sin}}\alpha }&{\cos \alpha }&0\\
0&0&1
\end{array}} \right],{{\bf{R}}_\beta } = \left[ {\begin{array}{*{20}{c}}
1&0&0\\
0&{\cos \beta }&{{\rm{ - sin}}\beta }\\
0&{{\rm{sin}}\beta }&{\cos \beta }
\end{array}} \right],{{\bf{R}}_\gamma } = \left[ {\begin{array}{*{20}{c}}
{\cos \gamma }&0&{{\rm{sin}}\gamma }\\
0&1&0\\
{ - {\rm{sin}}\gamma }&0&{\cos \gamma }
\end{array}} \right].
\end{align}

Let ${\bf{R = }}{{\bf{R}}_\alpha }{{\bf{R}}_\beta }{{\bf{R}}_\gamma }$ denote the rotated matrix.
 After the rotation,  we obtain the UE coordinate system ${X_1}$-${Y_1}$-${Z_1}$. To coincide with the UE coordinate system, we assume that the initial corresponding UE normal vector is ${{\bf{n}}_{{\rm{UE}}}} = {\left[ {0,0,1} \right]^T}$. It is clear that ${{\bf{n}}_{{\rm{UE}}}} = {{\bf{R}}^{ - 1}}{{\widehat{\bf n}}_{{\rm{UE}}}}$
   based on the coordinate transformation theory.

%  For analyzing OAR model briefly, we

  \subsection{OAR Model}
\begin{figure}[htbp]
\centering
\includegraphics[width=6.0cm]{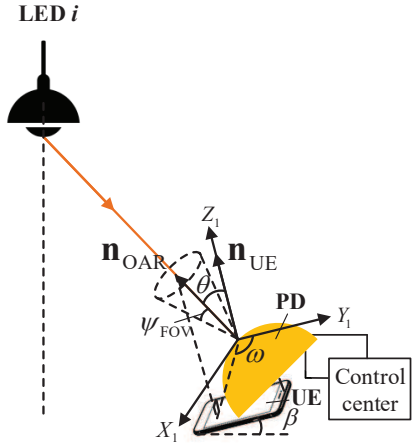}
\caption{ The directional geometry of OAR.} \label{OAR_fig}
\end{figure}
The alignment of the PD and the LED can be mathematically described via the OAR model.  Since the PD is adjustable, its orientation can be different from the UE orientation. Let ${{\bf{n}}_{{\rm{OAR}}}}\left( {\theta ,\omega } \right)$ denote  the normalized PD orientation vector in the UE coordinate ${X_1}$-${Y_1}$-${Z_1}$, where   the ${X_1}$-${Y_1}$ plane is parallel to the UE plane, and the $Z_1$ axis is perpendicular to the UE plane.
As shown in Fig. \ref{OAR_fig}, the orientation vector    ${{\bf{n}}_{{\rm{OAR}}}}\left( {\theta ,\omega } \right)$ can be expressed as
  \begin{align}
{{\bf{n}}_{{\rm{OAR}}}}\left( {\theta ,\omega } \right) =  {\left[ {{\rm{sin}}\theta \cos \omega ,{\rm{sin}}\theta {\rm{sin}}\omega ,\cos \theta } \right]^T}, \label{exp_nlen}
\end{align}
where $\theta$ is the  elevation angle between   ${{\bf{n}}_{{\rm{OAR}}}}\left( {\theta ,\omega } \right)$ and the ${Z_1}$ axis, and
  $\omega$ is the azimuth    angle between the projection of ${{\bf{n}}_{{\rm{OAR}}}}\left( {\theta ,\omega } \right)$ in the ${X_1}$-${Y_1}$ plane and the ${Y_1}$ axis.

By controlling $\theta$ and $\omega$, PD can be aligned with the LED as much as possible. Generally, physical  schemes to realize OAR are a mechanical   control receiver (MCR) and a tunable liquid-crystal receiver (TLR), which are discussed next.
\begin{itemize}
 \item  Mechanical   Control Receiver:
 The PD orientation  is adjusted by micro
electromechanical systems. More specifically,  MCR can freely adjust the PD rotation by a sophisticated mechanical module, and thus the PD direction ${{\bf{n}}_{{\rm{OAR}}}}\left( {\theta ,\omega } \right)$ can be aligned with the light incident direction
to some extent.   As shown in Fig.  \ref{OAR_fig},   the angle between the UE   and the horizontal plane is $\beta$, and after adjusting the  PD orientation  by MCR,
the PD orientation vector ${{\bf{n}}_{{\rm{OAR}}}}\left( {\theta ,\omega } \right)$ can be aligned with the LED.
 Consequently, when the user is moving, the PD of the receiver can be aimed at  the strongest incident angle of the light in real time.

\item Tunable Liquid-crystal Receiver:
A TLR   can
 effectively manipulate the angle of incident light at the PD by using the external electrical facility\cite{Alain_IWC_2021}.
Specifically, TLR is
a synthesized material composed of  dielectric structures and liquid-crystal cells,
 which are used to manipulate the light propagation
 in unusual ways compared to classical optical devices.
The  liquid-crystal cell is capable of
 realizing the adjustable refractive index,
 which exactly affects the direction and intensity of refracted light \cite{Alan_Sci_2018}, \cite{Sun_Nature_2019}.
  As shown in Fig.  \ref{OAR_fig},
  TLR can change its orientation  ${{\bf{n}}_{{\rm{OAR}}}}\left( {\theta ,\omega } \right)$ to
  steer    the LED.
\end{itemize}

With orientation  adjustable  schemes MCR or TLR, the   orientation of
PDs may perfectly align with the LED  when the UE orientation is fixed.
However, when the UE is moving, the UE orientation may change and can be random. Thus, the   orientation of
PDs may not accurately and timely align with the LED, because it takes time to adjust the orientation.
The fixed and random UE orientation cases will be further discussed in Sections III and IV, respectively.

\subsection{Blockage Model}
\begin{figure}[!htp]
\centering
\subfigure[System model.]{
\begin{minipage}[t]{0.25\linewidth}
\centering
\hspace*{-2.2cm}
\includegraphics[width=5cm]{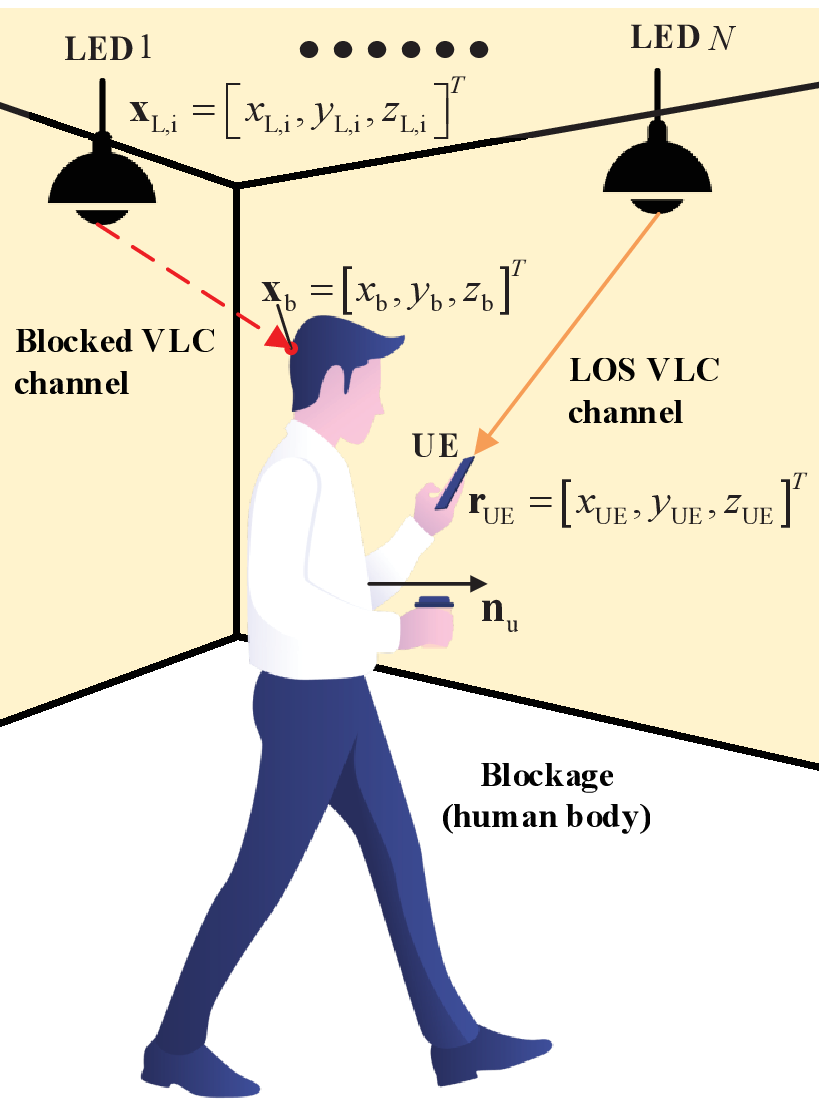}
%\caption{fig1}
\end{minipage}%
}%
\quad
\subfigure[Blockage model.]{
\begin{minipage}[t]{0.25\linewidth}
\centering
\hspace*{0cm}
\includegraphics[width=4.5cm]{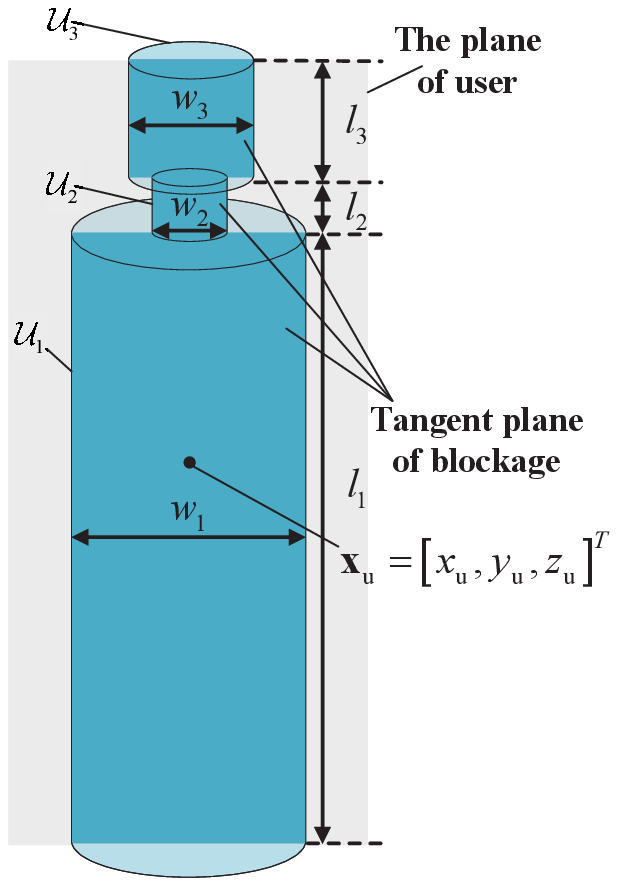}
%\caption{fig2}
\end{minipage}%
}%
\caption{ The schematic of   the blockage model of mobile UE.} \label{Blockage_fig}
\end{figure}

Due to the UE mobility, the LOS link may suffer from blockage, which would significantly influence the channel power gain. In practice, the VLC is vulnerable to blocking, especially by the mobile user {itself \cite{Tang_ICT_2021, Aboagye_ICT_2021}}. As shown in Fig. \ref{Blockage_fig} (a), let ${{\bf{r}}_{{\rm{UE}}}} = {\left[ {{x_{{\rm{UE}}}},{y_{{\rm{UE}}}},{z_{{\rm{UE}}}}} \right]^T}$ and ${{\bf{x}}_{{\rm{L}},i}} = {\left[ {{x_{{\rm{L}},i}},{y_{{\rm{L}},i}},{z_{{\rm{L}},i}}} \right]^T}$ denote the UE position  and the $i$th LED, respectively. Moreover,
${{\bf{n}}_{\rm{u}}}$ is defined as the normal vector of the human's forward direction, and ${{\bf{x}}_{\rm{b}}} = {\left[ {{x_{\rm{b}}},{y_{\rm{b}}},{z_{\rm{b}}}} \right]^T}$
is defined as an intersecting point of the blocked incident light  and the human body.

To characterize the blockage model,
    we
    model  the    head,    neck and body  of the person as three cylinders  ${{\cal U}_1}$, ${{\cal U}_2}$ and ${{\cal U}_3}$, respectively, as shown in Fig. \ref{Blockage_fig} (b).
  Specifically, let $w_i$ and $l_i$ denote the diameter and the height  of the cylinder  ${{\cal U}_i}$, respectively, where $i = 1,2$  and $3$. Moreover, let ${{\bf{x}}_{\rm{u}}} = {\left[ {{x_{\rm{u}}},{y_{\rm{u}}},{z_{\rm{u}}}} \right]^T}$ denote the center of gravity  position of the person. The  area  of  ${{\cal U}_1}$  can be expressed as
  \begin{align}
&{{\cal U}_1} \buildrel \Delta \over = \left\{ {{{\bf{x}}_0}\left| \begin{array}{l}
{\left( {{x_0} - {x_{\rm{u}}}} \right)^2} + {\left( {{y_0} - {y_{\rm{u}}}} \right)^2} \le {\left( {\frac{{{w_1}}}{2}} \right)^2},\\
0 \le {\bf{x}}_{\rm{0}}^T{{\bf{e}}_3} \le {l_1}
\end{array} \right.} \right\},\label{blockage_set_a}
\end{align}
  where   ${{\bf{e}}_3} \buildrel \Delta \over = {\left[ {0,0,1} \right]^T}$.
 Similarly, the  areas  of  ${{\cal U}_2}$ and ${{\cal U}_3}$    are respectively given as
 \begin{align}
 &{{\cal U}_2} \buildrel \Delta \over = \left\{ {{{\bf{x}}_0}\left| \begin{array}{l}
{\left( {{x_0} - {x_{\rm{u}}}} \right)^2} + {\left( {{y_0} - {y_{\rm{u}}}} \right)^2} \le {\left( {\frac{{{w_2}}}{2}} \right)^2},\\
{l_1} \le {\bf{x}}_{0}^T{{\bf{e}}_3} \le \left( {{l_1} + {l_2}} \right)
\end{array} \right.} \right\},\\
&{{\cal U}_3} \buildrel \Delta \over = \left\{ {{{\bf{x}}_0}\left| \begin{array}{l}
{\left( {{x_0} - {x_{\rm{u}}}} \right)^2} + {\left( {{y_0} - {y_{\rm{u}}}} \right)^2} \le {\left( {\frac{{{w_3}}}{2}} \right)^2},\\
\left( {{l_1} + {l_2}} \right) \le {\bf{x}}_{0}^T{{\bf{e}}_3} \le \left( {{l_1} + {l_2} + {l_3}} \right)
\end{array} \right.} \right\}.\label{blockage_set_c}
\end{align}

Since we focus on the LOS channel of the VLC link, the blockage may occur when the following two conditions happen simultaneously:  the direction of the incident light points to the back of the human body, since the UE is usually held in front of the human body; and the human body is at the LOS between the LED and the UE.
These two conditions can be mathematically described as  follows

\textbf{Condition ${{\rm{I}}}$:} When the direction of the incident light points to the back of the human, the angle between  the human forward direction vector ${{\bf{n}}_{\rm{u}}}$ and incident light vector $\left( {{{\bf{x}}_{{\rm{L}},i}} - {{\bf{r}}_{{\rm{UE}}}}} \right)$ is an obtuse angle, i.e.,
\begin{align}\label{direction_constrain}
{\bf{n}}_{\rm{u}}^T\left( {{{\bf{x}}_{{\rm{L}},i}} - {{\bf{r}}_{{\rm{UE}}}}} \right) \le 0.
\end{align}

\textbf{Condition ${\rm{II}}$:} There is an intersection point ${\bf x}_{\rm b}$ between the incident light line and the   human body.  Specifically, the line equation between the $i$th LED and the UE is given as
\begin{align}\label{constraints_blo}
\frac{{{x_{\rm{b}}} - {x_{{\rm{UE}}}}}}{{{x_{{\rm{L,i}}}} - {y_{{\rm{UE}}}}}} = \frac{{{y_{\rm{b}}} - {y_{{\rm{UE}}}}}}{{{y_{{\rm{L,i}}}} - {y_{{\rm{UE}}}}}} = \frac{{{z_{\rm{b}}} - {z_{{\rm{UE}}}}}}{{{z_{{\rm{L,i}}}} - {z_{{\rm{UE}}}}}}.
\end{align}
Using a tangent plane of
blockage as the constrained blockage plane, as shown in Fig.
1(b), which can be achieved by cutting the cylinder model. Specifically,
 the tangent blockage plane is given as
\begin{align}\label{blockage_set}
\left\{ \begin{array}{l}
{{\bf{x}}_{\rm{b}}} \in \left\{ {{{\cal U}_1} \cup {{\cal U}_2} \cup {{\cal U}_3}} \right\},\\
{\bf{n}}_{\rm{u}}^T\left( {{{\bf{x}}_{\rm{b}}} - {{\bf{x}}_{\rm{u}}}} \right) = 0.
\end{array} \right.
\end{align}
Thus, we have obtained intersection point ${\bf x}_{\rm b}$ by jointly solving Equations  \eqref{constraints_blo} and \eqref{blockage_set}.

Therefore, when  the Conditions ${\rm{I}}$ and ${\rm{II}}$ are met simultaneously, the incident light is blocked by the human body. We use an indicator parameter  ${u_i}$  to describe whether the LOS link  between the $i$th LED and UE is blocked or not, i.e., ${u_i}= 1$ means LOS link is blocked, while ${u_i}= 0$ means LOS link is not blocked.

\subsection{ Channel Gain and Achievable Rate Expressions}

According to the Lambertian model,  the channel gain  ${g_i}$ for the $i$th LED can be expressed as
\begin{align}\label{channel_gain}
{g_i} = \frac{{u_i}{\left( {m + 1} \right){A_{\rm{r}}}}}{{2\pi d_i^2}}{\cos ^m}\left( {{\phi _i}} \right){\Gamma _i}\left( {{\varphi _i},{\psi _{{\rm{FOV}}}}} \right)\cos \left( {{\varphi _i}} \right),
\end{align}
where  ${\Gamma _i}\left( {{\varphi _i},{\psi _{{\rm{FOV}}}}} \right)=1$ when   $\left| {{\varphi _i}} \right| \le {\psi _{{\rm{FOV}}}}$ and otherwise  ${\Gamma _i}\left( {{\varphi _i},{\psi _{{\rm{FOV}}}}} \right)=0$.  The parameter    ${{\psi _{{\text{FOV}}}}}$ is
  the field of vision   (FOV) of PD, $m = {{ - \ln 2} \mathord{\left/
 {\vphantom {{ - \ln 2} {\ln \left( {\cos \left( {{\phi _{{1 \mathord{\left/
 {\vphantom {1 2}} \right.
 \kern-\nulldelimiterspace} 2}}}} \right)} \right)}}} \right.
 \kern-\nulldelimiterspace} {\ln \left( {\cos \left( {{\phi _{{1 \mathord{\left/
 {\vphantom {1 2}} \right.
 \kern-\nulldelimiterspace} 2}}}} \right)} \right)}}$ is the
Lambertian index of the LED,  ${\phi _{{1 \mathord{\left/
 {\vphantom {1 2}} \right.
 \kern-\nulldelimiterspace} 2}}}$ is the semi-angle,
  ${d_i}$ is the distance between the $i$th LED
 and the  PD,  ${{A_{\rm{r}}}}$ is the effective area of the PD,  ${{\phi}}_i$  is the    irradiance angle, and ${\varphi _i}$  is the incidence angle.

 Based on the OAR  and the blockage models, we   derive  the channel gain expression as a function of the variables  ${\theta ,\omega }$ in the UE coordinate system.
  Let ${{{\tilde{ \bf x}}}_{{\rm{L}},i}}$ and ${{{\tilde{\bf r}}}_{{\rm{UE}}}}$, respectively, denote   the locations  of the $i$th LED and UE in the UE coordinate system ${X_1}$-${Y_1}$-${Z_1}$, which are given as
\begin{align}\label{normal_len_rot}
{{{\tilde{ \bf x}}}_{{\rm{L}},i}} = {{\bf{R}}^{ - 1}}{{\bf{x}}_{{\rm{L}},i}},{{{\tilde{\bf r}}}_{{\rm{UE}}}} = {{\bf{R}}^{ - 1}}{{\bf{r}}_{{\rm{UE}}}}.
\end{align}

Therefore, the terms $\cos \left( {{\phi _i}} \right)$ and $\cos \left( {{\varphi _i}} \right)$ in \eqref{channel_gain} can be,  respectively, rewritten as
  \begin{subequations}\label{gain_convert}
\begin{align}
&\cos \left( {{\phi _i}} \right) = \frac{{{z_{{\rm{L}},i}} - {z_{{\rm{UE}}}}}}{{{d_i}}},\label{gain_convert_a}\\
&\cos \left( {{\varphi _i}} \right) = \frac{{{{\left( {{{\tilde{ \bf x}}_{{\rm{L}},i}} - {{\tilde{\bf r}}_{{\rm{UE}}}}} \right)}^T}{{\bf{n}}_{{\rm{OAR}}}}\left( {\theta ,\omega } \right)}}{{{d_i}}}\label{gain_convert_b}.
\end{align}
\end{subequations}

Substituting \eqref{gain_convert} into  \eqref{channel_gain},  the channel gain ${g_i}\left( {\theta ,\omega } \right)$ between the $i$th LED and the UE can be re-expressed as
\begin{align}\label{gain}
{g_i}\left( {\theta ,\omega } \right) = {\lambda _i}{{\bf{d}}_i}^T{{\bf{n}}_{{\rm{OAR}}}}\left( {\theta ,\omega } \right){{\Gamma _i}\left( {{{\bf{n}}_{{\rm{OAR}}}}\left( {\theta ,\omega } \right)} \right)},
\end{align}
where ${\lambda _i} = \frac{{{u_i}{A_{\rm{r}}}\left( {m + 1} \right){{\left( {{z_{{\rm{L}},i}} - {z_{{\rm{UE}}}}} \right)}^m}}}{{2\pi d_i^{m + 3}}}$, ${{\bf{d}}_i}  = {{{\tilde{ \bf x}}_{{\rm{L}},i}} - {{\tilde{ \bf r}}_{{\rm{UE}}}}}$ is the incident vector between the $i$th LED and the UE in the UE coordinate system, and ${\Gamma _i}\left( {{{\bf{n}}_{{\rm{OAR}}}}\left( {\theta ,\omega } \right)} \right) $ is the indicator function of ${{\bf{n}}_{{\rm{OAR}}}}\left( {\theta ,\omega } \right)$ given as
\begin{align}
{{\Gamma _i}\left( {{{\bf{n}}_{{\rm{OAR}}}}\left( {\theta ,\omega } \right)} \right) = \left\{ {\begin{array}{*{20}{l}}
{1,}&{\frac{{{\bf{d}}_i^T{{\bf{n}}_{{\rm{OAR}}}}\left( {\theta ,\omega } \right)}}{{{d_i}}} \ge \cos \left( {{\psi _{{\rm{FOV}}}}} \right)},\\
{0,}&{\;\;\;\;\;\;{\rm{otherwise}}.}
\end{array}} \right.}
\end{align}

% Then, we assume that the input signal $s$ follows the  ABG
%distribution  \cite{Ma_ITC_2019}
%and the achievable rate ${R\left( {\theta ,\omega ,{\bf{p}}} \right)}$ of the VLC system  is given as
%\begin{align}\label{rate}
%{R\left( {\theta ,\omega ,{\bf{p}}} \right)} = B{\log _2}\left( {1 + \frac{{{{\left| {{{{{\bf{g}}\left( {\theta ,\omega } \right)}}^T}{\bf{p}}} \right|}^2}{e^{1 + 2\left( {\alpha_0  + \gamma_0 \varepsilon_0 } \right)}}}}{{2\pi B{\sigma ^2}}}} \right),
%\end{align}
%where $B$ denotes the bandwidth of the VLC system,  $\alpha_0$, $\gamma_0$ and $\varepsilon_0$ are calculated based on the equation given in \cite{Ma_ITC_2019}.

Since the   channel capacity of VLC channels is unknown, we adopt the  ABG lower bound  with close-form expression {\cite{Ruixin_JLT_2017}} analyze the mobile VLC system, i.e.,
\begin{align}\label{rate}
{R\left( {\theta ,\omega ,{\bf{p}}} \right)} = B{\log _2}\left( {1 + \frac{{{{\left| {{{{{\bf{g}}\left( {\theta ,\omega } \right)}}^T}{\bf{p}}} \right|}^2}{e^{1 + 2\left( {\alpha_0  + \gamma_0 \varepsilon } \right)}}}}{{2\pi B{\sigma ^2}}}} \right),
\end{align}
where $B$ denotes the bandwidth of the VLC system. Moreover, the parameters $\alpha_0 $, $\beta_0 $ and $\gamma_0$ are determined by the inputs constraints  $\left| s \right| \le A$, $\mathbb{E}\left\{ s \right\} = 0$ and $\mathbb{E}\left\{ {{s^2}} \right\} = \varepsilon $ as follows
\begin{subequations}
\begin{align}
 &T\left( A \right) - T\left( { - A} \right) = {e^{1+\alpha_0 }},\label{f_eq_a}\\
&\beta_0 \left( {{e^{A\left( {\beta_0  - \gamma_0 A} \right)}} - {e^{ - A\left( {\beta_0  + \gamma_0 A} \right)}} - {e^{1+\alpha_0 }}} \right) = 0,\label{f_eq_b}\\
&{e^{A\left( {\beta_0  - \gamma_0 A} \right)}}\left( {\left( {\beta_0  - 2\gamma_0 A} \right){e^{ - 2A\beta_0 }} - \beta_0  - 2\gamma_0 A} \right) + \left( {{\beta_0 ^2} + 2\gamma_0 } \right){e^{1+\alpha_0 }} = 4{\gamma_0 ^2}\varepsilon {e^{1 + \alpha_0 }},\label{f_eq_c}
 \end{align}
\end{subequations}
where
$T\left( x \right) \buildrel \Delta \over = \frac{{\sqrt \pi  }}{{2\sqrt {{\gamma _0}} }}{e^{\frac{{\beta _0^2}}{{4{\gamma _0}}}}}{\rm{erf}}\left( {\frac{{{\beta _0} + 2{\gamma _0}x}}{{2\sqrt {{\gamma _0}} }}} \right)$.

\section{Joint  BO Scheme Design  for Fixed UE Orientation}
To begin with, we consider the case of  fixed UE orientation, i.e.,  the normal orientation vector ${{\bf{n}}_{{\rm{UE}}}}$ of UE is fixed with certain angles,
and the orientations of PD can be adjusted to align with the LED.
Based on the OAR's orientation analysis in the previous section, we further jointly optimize BO, i.e., the elevation angle $\theta$ and azimuth angle $\omega$, and beamformer  ${\bf{p}}$,   to minimize the total transmit power of LEDs, while satisfying both the orientation angle and rate constraints.
 Mathematically,  the joint beamforming and orientation optimization problem can be formulated as
\begin{subequations}\label{power_prob_o}
\begin{align}
\mathop {\min }\limits_{\theta ,\omega ,{\bf{p}}} ~&\varepsilon {\left\| {\bf{p}} \right\|^2}\label{power_prob_o_a}\\
{\rm{s}}{\rm{.t}}{\rm{.}}~&{R\left( {\theta ,\omega ,{\bf{p}}} \right)} \ge {\bar R},\label{power_prob_o_b}\\
&0 \le \theta  \le  \bar \theta  ,\label{power_prob_o_c}\\
& - \pi  \le \omega  \le \pi ,\label{power_prob_o_d}\\
&\sqrt {{p_i}} A \le {I_{{\rm{DC}}}},\forall i \in {\cal N}\label{power_prob_o_e},
\end{align}
\end{subequations}
where  ${\bar R}$ is the minimum rate requirement of the UE, and
  $\bar \theta  \in \left( {0,\left. {\frac{\pi }{2}} \right]} \right.$ is the elevation angle $\theta$ threshold.
  Since   the elevation angle $\theta$ and azimuth angle $\omega$ are coupled  in \eqref{power_prob_o_b},  problem \eqref{power_prob_o} is non-convex
and computationally intractable.
To address this challenge,
we  first reformulate problem \eqref{power_prob_o} into a more concise form by  merging the  optimization variables.
Recall that ${\left\| {{{\bf{n}}_{{\rm{OAR}}}}} \right\|^2} = 1$, since  the orientation vector ${{\bf{n}}_{{\rm{OAR}}}}\left( {\theta ,\omega } \right)$ is a normal vector.
 Given that  $\cos  \theta $ is a monotonically decreasing function for $0 \le \theta  \le  \bar \theta$ and  ${\rm{sin}}\omega$ is a monotonically increasing function for  $- \pi  \le \omega  \le \pi$,  ${{\bf{n}}_{{\rm{OAR}}}}\left( {\theta ,\omega } \right)$ is a one-to-one mapping of variables  $\left( {\theta ,\omega } \right)$.
  Then, since $0 \le \theta  \le  \bar \theta$,   we have
${\cos ^2}\bar \theta  \le \cos \theta  \le 1$.
 Thus, by projecting the orientation vector ${{\bf{n}}_{{\rm{OAR}}}}\left( {\theta ,\omega } \right)$  into the ${X_1}$-${Y_1}$ plane, we have
\begin{align}
&{\left( {{\rm{sin}}\theta \cos \omega } \right)^2} + {\left( {{\rm{sin}}\theta {\rm{sin}}\omega } \right)^2} \le {\sin ^2}\bar \theta.
\end{align}

   Furthermore, constraints \eqref{power_prob_o_c} and \eqref{power_prob_o_d}   can be   equivalently rewritten as
\begin{subequations}\label{conatrain_nlen}
\begin{align}
&{\left\| {\left( {{{\bf{e}}_1} + {{\bf{e}}_2}} \right) \odot {{\bf{n}}_{{\rm{OAR}}}}} \right\|^{\rm{2}}} \le {\sin ^2}\bar \theta ,\\
&\cos \bar \theta  \le {{\bf{n}}_{{\rm{OAR}}}^T}{{\bf{e}}_3} \le {\rm{1}},
\end{align}
\end{subequations}
where ${{\bf{e}}_i}$ is a unit vector with the $i$-th element equal to $1$,   $i=1,2$ and $3$.
  Therefore, based on the orientation vector ${{\bf{n}}_{{\rm{OAR}}}}$, problem \eqref{power_prob_o} can be equivalently reformulated as
\begin{subequations}\label{power_prob_oo}
\begin{align}
\mathop {\min }\limits_{{{\bf{n}}_{{\rm{OAR}}}} ,{\bf{p}}} ~&\varepsilon {\left\| {\bf{p}} \right\|^2}\label{power_prob_oo_a}\\
{\rm{s}}{\rm{.t}}{\rm{.}}~&B{\log _2}\left( {1 + \frac{{{{\left| {{{{{{{\bf{g}}\left( {{{\bf{n}}_{{\rm{OAR}}}} } \right)}}}}^T}{\bf{p}}} \right|}^2}{e^{1 + 2\left( {\alpha_0  + \gamma_0 \varepsilon_0 } \right)}}}}{{2\pi B{\sigma ^2}}}} \right) \ge {\bar R},\label{power_prob_oo_b}\\
&{\left\| {\left( {{{\bf{e}}_1} + {{\bf{e}}_2}} \right) \odot {{\bf{n}}_{{\rm{OAR}}}}} \right\|^{\rm{2}}} \le {\sin ^2}\bar \theta ,\label{power_prob_oo_c}\\
&\cos \bar \theta  \le {{\bf{n}}_{{\rm{OAR}}}^T}{{\bf{e}}_3} \le {\rm{1}},\label{power_prob_oo_d}\\
&{\left\| {{{\bf{n}}_{{\rm{OAR}}}}} \right\|^2} = 1\label{power_prob_oo_e},\\
&\sqrt {{p_i}} A \le {I_{{\rm{DC}}}},\forall i \in {\cal N}\label{power_prob_oo_f}.
\end{align}
\end{subequations}

Note that, due to the coupling of  the beamforming  vector ${\bf{p}}$ and the orientation vector ${{\bf{n}}_{{\rm{OAR}}}}$,
 problem \eqref{power_prob_oo} is still non-convex and intractable.
 To make this  complicated problem tractable, we first decouple the beamforming  vector ${\bf{p}}$ and the orientation vector ${{\bf{n}}_{{\rm{OAR}}}}$ by decomposing  problem \eqref{power_prob_oo} into two sub-problems,
  i.e., a beamforming design subproblem and a PD orientation subproblem.
 Specifically, we propose an alternating optimization and projection algorithm to handle  problem \eqref{power_prob_oo},
 in which the {transmit} beamforming vector ${\bf{p}}$ and the PD orientation vector ${{\bf{n}}_{{\rm{OAR}}}}$
  are  alternately optimized.

\subsection{Beamforming Design Subproblem}
For a given PD orientation  ${{\bf{n}}_{{\rm{OAR}}}}$,  the beamforming design subproblem of \eqref{power_prob_o} is given as \begin{subequations}\label{power_prob1_o}
\begin{align}
\mathop {\min }\limits_{{\bf{p}}} ~&\varepsilon {\left\| {\bf{p}} \right\|^2}\label{power_prob1_o_a}\\
{\rm{s}}{\rm{.t}}{\rm{.}}~&B{\log _2}\left( {1 + \frac{{{{\left| {{{{{{{\bf{g}}\left( {{{\bf{n}}_{{\rm{OAR}}}} } \right)}}}}^T}{\bf{p}}} \right|}^2}{e^{1 + 2\left( {\alpha_0  + \gamma_0 \varepsilon_0 } \right)}}}}{{2\pi B{\sigma ^2}}}} \right) \ge {\bar R},\label{power_prob1_o_b}\\
&\sqrt {{p_i}} A \le {I_{{\rm{DC}}}},\forall i \in {\cal N}\label{power_prob1_o_c},
\end{align}
\end{subequations}
which  is also non-convex.
To address the non-convexity issue, we apply the SDR technique to relax problem \eqref{power_prob1_o}. Specifically,  by using the following relationship
\begin{align}
{\bf{P}} = {\bf{p}}{{\bf{p}}^T} \Leftrightarrow {\rm{rank}}\left( {\bf{P}} \right) = 1,{\bf{P}}\succeq{\bf{0}},
\end{align}
and neglecting the non-convex rank-$1$ constraint, problem \eqref{power_prob1_o} can be reformulated as
\begin{subequations}\label{power_prob1}
\begin{align}
 \mathop {\min }\limits_{{\bf{P}}}  &~\varepsilon{\rm{Tr}}\left( {\bf{P}} \right)\\
{\rm{s}}{\rm{.t}}{\rm{.}}~&{\rm{Tr}}\left( {{\bf{Pg}}\left( {{{\bf{n}}_{{\rm{OAR}}}} } \right){\bf{g}}{{\left( {{{\bf{n}}_{{\rm{OAR}}}}} \right)}^T}} \right) \ge {c_1},\\
&{\rm{Tr}}\left( {{\bf{P}}{{\bf{e}}_i}{\bf{e}}_i^T} \right) \le \frac{{I_{{\rm{DC}}}^2}}{A},\forall i \in {\cal N},\\
&{\bf{P}}\succeq{\bf{0}},
\end{align}
\end{subequations}
where ${c_1} = \left( {{2^{\frac{{\bar R}}{B}}} - 1} \right)\frac{{2\pi B{\sigma ^2}}}{{{e^{1 + 2\left( {\alpha_0  + \gamma_0 \varepsilon} \right)}}}}$.
Problem \eqref{power_prob1} is a convex semidefinite program (SDP), and the optimal beamforming vector ${{\bf{P}}_{\rm{o}}}$ can be obtained by interior-point algorithms \cite{Ye_1997}\cite{Grant_2009}. The computation complexity of    \eqref{power_prob1} is ${\mathcal O}\left( {{{\left( {N + 2} \right)}^4}{N^{{1 \mathord{\left/
 {\vphantom {1 2}} \right.
 \kern-\nulldelimiterspace} 2}}}\log \left( {{1 \mathord{\left/
 {\vphantom {1 {\zeta}_1 }} \right.
 \kern-\nulldelimiterspace} {\zeta}_1 }} \right)} \right)$, where ${\zeta}_1 > 0$ is the solution accuracy {\cite{Luo_ISPM_2010}}. Note that if ${\rm{rank}}\left( {{\bf{P}}_{\rm{o}}} \right) = 1$, the optimal beamforming vector ${{\bf{p}}_{{\rm{o}}}}$ of problem \eqref{power_prob1_o} can be obtained by eigenvalue decomposition.  Due to SDR, the case ${\rm{rank}}\left( {{\bf{P}}_{\rm{o}}} \right) > 1$ may also occur.  In this case, we can use the Gaussian
randomization procedure to generate a high-quality feasible beamformer vector ${{\bf{p}}_{{\rm{o}}}}$ \cite{Luo_ISPM_2010}.

\subsection{PD Orientation  Subproblem}
For a given beamformer vector ${{\bf{p}}_{{\rm{o}}}}$,  the PD orientation optimization subproblem is given by
 \begin{subequations}\label{power_prob2_o}
\begin{align}
\mathop {\min }\limits_{{{\bf{n}}_{{\rm{OAR}}}} } ~&\varepsilon {\left\| {\bf{p_{\rm{o}}}} \right\|^2}\label{power_prob2_o_a}\\
{\rm{s}}{\rm{.t}}{\rm{.}}~&B{\log _2}\left( {1 + \frac{{{{\left| {{{{{{{\bf{g}}\left( {{{\bf{n}}_{{\rm{OAR}}}} } \right)}}}}^T}{{\bf{p}}_{\rm{o}}}} \right|}^2}{e^{1 + 2\left( {\alpha_0  + \gamma_0 \varepsilon_0 } \right)}}}}{{2\pi B{\sigma ^2}}}} \right) \ge {\bar R},\label{power_prob2_o_b}\\
&{\left\| {\left( {{{\bf{e}}_1} + {{\bf{e}}_2}} \right) \odot {{\bf{n}}_{{\rm{OAR}}}}} \right\|^{\rm{2}}} \le {\sin ^2}\bar \theta ,\label{power_prob2_o_c}\\
&\cos \bar \theta  \le {{\bf{n}}_{{\rm{OAR}}}^T}{{\bf{e}}_3} \le {\rm{1}},\label{power_prob2_o_d}\\
&{\left\| {{{\bf{n}}_{{\rm{OAR}}}}} \right\|^2} = 1.\label{power_prob2_o_e}
\end{align}
\end{subequations}

Note  that problem \eqref{power_prob2_o} is an  optimization problem of finding feasible solutions satisfying both the orientation constraints  and the minimum rate   constraint. However, there are many feasible solutions satisfying constraints \eqref{power_prob2_o_b},\eqref{power_prob2_o_c} and \eqref{power_prob2_o_d}.
Moreover,  the objective function \eqref{power_prob2_o_a}      decreases with the power of beamforming vector ${\left\| {{{\bf{p}}_{\rm{o}}}} \right\|^2}$, and is independent of  ${{\bf{n}}_{{\rm{OAR}}}}$.
Thus, we  optimize the orientation vector ${{\bf{n}}_{{\rm{OAR}}}}$ to minimize the transmission power with the maximum  ${{\bf{g}}{{\left( {{{\bf{n}}_{{\rm{OAR}}}} } \right)}^T}{{\bf{p}}_{\rm{o}}}}$.  Then, the PD orientation subproblem \eqref{power_prob2_o} can be reformulated as
\begin{align}
\mathop {\max }\limits_{{{\bf{n}}_{{\rm{OAR}}}} } ~&{{\bf{g}}{{\left( {{{\bf{n}}_{{\rm{OAR}}}} } \right)}^T}{{\bf{p}}_{\rm{o}}}}\label{power_prob2_oo_a}\\
{\rm{s}}{\rm{.t}}{\rm{.}}~&\eqref{power_prob2_o_b},\eqref{power_prob2_o_c},
\eqref{power_prob2_o_d},\eqref{power_prob2_o_e}\nonumber.
\end{align}

To address  subproblem  (27), the objective function of \eqref{power_prob2_oo_a} with PD orientation vector ${{\bf{n}}_{{\rm{OAR}}}}$  can be reformulated as
\begin{align}\label{objctive_func}
{\bf{g}}{\left( {{{\bf{n}}_{{\rm{OAR}}}} } \right)^T}{{\bf{p}}_{\rm{o}}} = \sum\limits_{i = 1}^N {{\lambda _i}{\sqrt {{p_{i}}} }{\bf{d}}_i^T{{\bf{n}}_{{\rm{OAR}}}}{\Gamma _i}\left( {{{\bf{n}}_{{\rm{OAR}}}}} \right)}.
\end{align}

Based on the expansion expression \eqref{objctive_func}, constraint \eqref{power_prob2_o_b} can   be rewritten as
\begin{align}\label{conatrain_rbar}
\sum\limits_{i = 1}^N {{\lambda _i}{\sqrt {{p_{i}}}}{\bf{d}}_i^T{{\bf{n}}_{{\rm{OAR}}}}{\Gamma _i}\left( {{{\bf{n}}_{{\rm{OAR}}}}} \right)}  \ge {c_2},
\end{align}
where ${c_2} = \sqrt {\left( {{2^{\frac{{\bar R}}{B}}} - 1} \right)\frac{{2\pi B{\sigma ^2}}}{{{e^{1 + 2\left( {\alpha_0  + \gamma_0 \varepsilon_0} \right)}}}}} $.
Furthermore,    we relax the equality constraint  ${\left\| {{{\bf{n}}_{{\rm{OAR}}}}} \right\|^2} = 1$ as ${\left\| {{{\bf{n}}_{{\rm{OAR}}}}} \right\|^2} \le 1$.
Therefore,  the PD orientation  subproblem {(27)} can be reformulated as
\begin{subequations}\label{rate_prob1_o}
\begin{align}
 \mathop {\max }\limits_{{{{{{\bf n}}}_{{\rm{OAR}}}}}}~&{\sum\limits_{i = 1}^N {{\lambda _i}{\sqrt {{p_{i}}}}{\bf{d}}_i^T{{\bf{n}}_{{\rm{OAR}}}}{{\Gamma _i}\left( {{{\bf{n}}_{{\rm{OAR}}}}} \right)}} }\label{rate_prob1_o_a}\\
{\rm{s}}{\rm{.t}}{\rm{.}}~&\sum\limits_{i = 1}^N {{\lambda _i}{\sqrt {{p_{i}}}}{\bf{d}}_i^T{{\bf{n}}_{{\rm{OAR}}}}{\Gamma _i}\left( {{{\bf{n}}_{{\rm{OAR}}}}} \right)}  \ge {c_2},\label{rate_prob1_o_b}\\
&{\left\| {\left( {{{\bf{e}}_1} + {{\bf{e}}_2}} \right) \odot {{\bf{n}}_{{\rm{OAR}}}}} \right\|^{\rm{2}}} \le {\sin ^2}\bar \theta ,\label{rate_prob1_o_c}\\
&\cos \bar \theta  \le {{\bf{n}}_{{\rm{OAR}}}^T}{{\bf{e}}_3} \le {\rm{1}},\label{rate_prob1_o_d}\\
&{\left\| {{{\bf{n}}_{{\rm{OAR}}}}} \right\|^2} \le 1.\label{rate_prob1_o_e}
\end{align}
\end{subequations}

Due to the integer and indicator function ${\Gamma _i}\left( {{{\bf{n}}_{{\rm{OAR}}}}} \right)$ in both objective function \eqref{rate_prob1_o_a} and constraint \eqref{rate_prob1_o_b},  problem \eqref{rate_prob1_o} is {a non-linear non-smooth problem \cite{Luenberger_1984}}, which is hard to solve .
To overcome the NP-hard challenging issue, we propose an  iterative optimization and projection method to handle problem \eqref{rate_prob1_o}.
To be more specific, during   the $k$th iteration,  we relax this    problem \eqref{rate_prob1_o}   as follows
\begin{subequations}\label{rate_prob1_2}
\begin{align}
 \mathop {\max }\limits_{{{{{{\bf n}}}_{{\rm{OAR}}}}}}~&{\sum\limits_{j \in {\mathcal M}} {{\lambda _j}{\sqrt {{p_{j}}}}{\bf{d}}_j^T{{{{\bf n}}}_{{\rm{OAR}}}}}}\\
{\rm{s}}{\rm{.t}}{\rm{.}}~&\sum\limits_{j \in {\mathcal M}} {{\lambda _j}{\sqrt {{p_{j}}}}{\bf{d}}_j^T{{\bf{n}}_{{\rm{OAR}}}}}  \ge {c_2},\\
&\eqref{rate_prob1_o_c},\eqref{rate_prob1_o_d},\eqref{rate_prob1_o_e},\nonumber
\end{align}
\end{subequations}
where ${\mathcal M} \buildrel \Delta \over = \left\{ {\forall j \in {\mathcal N}\left| {{\Gamma _j}\left( {\widetilde {\bf{n}}_{{\rm{OAR}}}^{\left[ m \right]}} \right)  \ne 0} \right.} \right\}$  describes the LEDs set. Then,  by applying the standard interior-point algorithm \cite{Ye_1997},
  problem \eqref{rate_prob1_2} can be efficiently
solved and the solution $\widetilde {\bf{n}}_{{\rm{OAR}}}^{\left[ m \right]}$  can be obtained.  Then, the LEDs set ${\mathcal M}$ is updated. Furthermore,   we decide the value of the indicator function ${\Gamma _i}\left( {{{\widetilde {\bf{n}}_{{\rm{OAR}}}^{\left[ 0 \right]}}}} \right)$ as follows. By updating the LEDs set  ${\mathcal M} $, ${\Gamma _i}\left( {{{\widetilde {\bf{n}}_{{\rm{OAR}}}^{\left[ 0 \right]}}}} \right)$ is {derived} by the following the principle: if $\frac{{{\bf{d}}_i^T{\widetilde {\bf{n}}_{{\rm{OAR}}}^{\left[ 0 \right]}}}}{{{d_i}}} < \cos \left( {{\psi _{{\rm{FOV}}}}} \right)$, ${\Gamma _i}\left( {{{\widetilde {\bf{n}}_{{\rm{OAR}}}^{\left[ 0 \right]}}}} \right)= 0$; otherwise, ${\Gamma _i}\left( {{{\widetilde {\bf{n}}_{{\rm{OAR}}}^{\left[ 0 \right]}}}} \right)= 1$.

Thus, the proposed PD orientation optimization and projection method is summarized in  Algorithm $1$.
  \renewcommand{\algorithmicrequire}{\textbf{Input:}}
\renewcommand{\algorithmicensure}{\textbf{Output:}}
\begin{algorithm}[H]
    \caption{Searching PD orientation}
    \label{dinkelbach_alg}
    \begin{algorithmic}[1]
         \State \textbf{Initialize:}   $ m = 0 $, a feasible beamforming vector ${{\bf{p}}_{{\rm{o}}}}$;
         \While{$\frac{{{\bf{d}}_j^T {\widetilde {\bf{n}}_{{\rm{OAR}}}^{\left[ m \right]}} }}{{{d_j}}} < \cos \left( {{\psi _{{\rm{FOV}}}}} \right)$}
%         \State $ m \leftarrow m+1 $;
         \State Calculating the problem \eqref{rate_prob1_2} by using $\widetilde {\bf{n}}_{{\rm{OAR}}}^{\left[ m \right]}$  and update set ${\mathcal M}$ ;
         \State $ m \leftarrow m+1 $;
          \EndWhile
          \State Obtain ${{\widetilde{\bf n}}}_{{\rm{OAR}}}^ *  \leftarrow {{\widetilde{\bf n}}}_{{\rm{OAR}}}^{\left[ m \right]}$.
%         \Ensure ${{\widetilde{\bf n}}}_{{\rm{len}}}^ *  \leftarrow {{\widetilde{\bf n}}}_{{\rm{len}}}^{\left[ m \right]}$
    \end{algorithmic}
\end{algorithm}

\subsection{AO algorithm for fixed UE orientation}

In the  previous section, we have optimized the beamforming  ${{\bf{P}}_{\rm{o}}}$ and PD orientation vector ${{\bf{n}}_{{\rm{OAR}}}}$. Then, we present the specific alternating optimization algorithm
to find the above vector in Algorithm 2.
Specifically, we initialize a given receiving orientation vector ${\bf{n}}_{{\rm{OAR}}}^{\left[ 0 \right]}$, and obtain a beamforming  ${\bf{P}}^{\left[ 1 \right]}$ by using the SDR technique. Then, by applying Algorithm 1, we update ${\bf{n}}_{{\rm{OAR}}}^{\left[ 1 \right]}$. By iteratively calculating ${\bf{n}}_{{\rm{OAR}}}^{\left[ k \right]}$ and ${\bf{P}}^{\left[ k \right]}$, the proposed Algorithm 2 will proceed until convergence. The details of algorithm are described as follows.
\renewcommand{\algorithmicrequire}{\textbf{Input:}}
\renewcommand{\algorithmicensure}{\textbf{Output:}}
\begin{algorithm}[H]
    \caption{  AO  Algorithm for Problem (22) }
    \label{dinkelbach_alg}
    \begin{algorithmic}[1]
%        \Require ${{\bf{P}}^{[0]}}$
%        \Ensure ${{\bf{P}}^ * },{{\bf{n}}_{{\rm{len}}}^ * }$
%        \State \textbf{Given} the convergence criterion ${\delta}  > 0$, iteration number $ k = 1 $;
        \State {\textbf{Initialize:} the convergence criterion ${\delta}  > 0$, iteration number $ k = 1 $;}
        \State  {Choose an orientation vector ${\bf{n}}_{{\rm{OAR}}}^{\left[ 0 \right]}$ satisfying $\sum\limits_{i = 1}^N {{g_i}\left( {{\bf{n}}_{{\rm{OAR}}}^{\left[ 0 \right]}} \right) \ge } \sqrt {\frac{{\left( {{2^{{{\bar R} \mathord{\left/
 {\vphantom {{\bar R} B}} \right.
 \kern-\nulldelimiterspace} B}}} - 1} \right)2\pi B{\sigma ^2}}}{{{e^{\left( {1 + 2\left( {{\alpha _0} + {\gamma _0}{\varepsilon _0}} \right)} \right)}}}}} \frac{A}{{{I_{{\rm{DC}}}}}}$;}
        \State Calculating ${\bf{P}}^{\left[ k \right]}$ based on \eqref{power_prob1};
        \State Calculating ${\bf{n}}_{{\rm{OAR}}}^{\left[ k \right]}$ based on Algorithm 1;
         \While{$\left\| {{\bf{n}}_{{\rm{OAR}}}^{\left[ k \right]} - {\bf{n}}_{{\rm{OAR}}}^{\left[ k-1 \right]}} \right\| \ge \delta $}
         \State $ k \leftarrow k+1 $;
         \State Go back to Step 3;
          \EndWhile
         \State Obtain ${{\bf{P}}^ * } \leftarrow {{\bf{P}}^{[k]}},{\bf{n}}_{{\rm{OAR}}}^ *  \leftarrow {\bf{n}}_{{\rm{OAR}}}^{\left[ k \right]}$.
    \end{algorithmic}
\end{algorithm}

 Ultimately, the above AO algorithm is devedoped to handle problem \eqref{power_prob_o}, which includes two key steps, i.e., solving the beamforming design subproblem in Step 3 and the PD orientation optimization subproblem in Step 4.
For obtaining beamforming vector ${\bf{P}}^{\left[ 1 \right]}$ and   PD orientation vector ${\bf{n}}_{{\rm{OAR}}}^{\left[ 0 \right]}$, we employ the interior-point algorithm.  In conclusion, the total computational complexity of the proposed AO algorithm is approximately  ${\cal O}\left( {\max {{\left\{ {N + 2,4} \right\}}^4}{\left({N+3}\right)^{{1 \mathord{\left/
 {\vphantom {1 2}} \right.
 \kern-\nulldelimiterspace} 2}}}\log \left( {{1 \mathord{\left/
 {\vphantom {1 {{\zeta _2}}}} \right.
 \kern-\nulldelimiterspace} {{\zeta _2}}}} \right)} \right)$, where ${\zeta _2} > 0$ is the solution accuracy {\cite{Luo_ISPM_2010}}.

\section{Robust Joint BO Optimization for   Random UE Orientation}

%Note that the  above proposed  joint     beamforming and orientation   optimization  based on
% the perfect UE orientation case  is available.
 In this section, we further investigate robust joint   BO scheme   for the case of random UE orientation, where
 the  orientation  ${{\bf{n}}_{{\rm{UE}}}}$ can be random within a certain range.

%\subsection{Achievable Rate with Angel Error}
\begin{figure}[H]
      \centering
	\includegraphics[width=4.2cm]{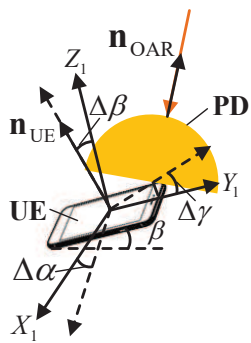}
      \vskip-0.2cm\centering
 \caption{ The  random UE orientation model. }
  \label{robust_fig} %% label for entire figure
\end{figure}
   Considering the random UE orientation scenario, as shown in Fig. \ref{robust_fig}, where
the UE rotation direction   is bounded  during information transmission.
Specifically, let
$\Delta \alpha  \in \left[ { - \bar \alpha ,\bar \alpha } \right)$, $\Delta \beta  \in \left[ { - \bar \beta ,\bar \beta } \right)$, and $\Delta \gamma  \in \left[ { - \bar \gamma ,\bar \gamma } \right)$ denote random angles, where $\bar \alpha$, $\bar \beta$ and $\bar \gamma$ denote the corresponding ranges of the random  angles.
Then, let ${{\bf{R}}_\Delta }  = {{\bf{R}}_{\Delta \alpha }}{{\bf{R}}_{\Delta \beta }}{{\bf{R}}_{\Delta \gamma }}$ denote the  corresponding UE rotated matrices, where
\begin{align}
{{\bf{R}}_{\Delta \alpha }} = \left[ {\begin{array}{*{20}{c}}
{\cos \Delta \alpha }&{{\rm{ - sin}}\Delta \alpha }&0\\
{{\rm{sin}}\Delta \alpha }&{\cos \Delta \alpha }&0\\
0&0&1
\end{array}} \right],
{{\bf{R}}_{\Delta \beta }} = \left[ {\begin{array}{*{20}{c}}
1&0&0\\
0&{\cos \Delta \beta }&{{\rm{ - sin}}\Delta \beta }\\
0&{{\rm{sin}}\Delta \beta }&{\cos \Delta \beta }
\end{array}} \right],
{{\bf{R}}_{\Delta \gamma }} = \left[ {\begin{array}{*{20}{c}}
{\cos \Delta \gamma }&0&{{\rm{sin}}\Delta \gamma }\\
0&1&0\\
{ - {\rm{sin}}\Delta \gamma }&0&{\cos \Delta \gamma }
\end{array}} \right],
\end{align}
where $\Delta \alpha  \in \left[ { - \bar \alpha ,\bar \alpha } \right)$ is the UE yaw random angle, while $\Delta \beta  \in \left[ { - \bar \beta ,\bar \beta } \right)$ and $\Delta \gamma  \in \left[ { - \bar \gamma ,\bar \gamma } \right)$ are the UE pitch and roll random angles, respectively.

Based on the above definition, the PD channel gain can be expressed as
\begin{align}
{g_i}\left( {{{\bf{n}}_{{\rm{OAR}}}}} \right)  = {\lambda _i}{\left( {{{\bf{x}}_{{\rm{L}},i}} - {{\bf{r}}_{{\rm{UE}}}}} \right)^T}{\left( {{\bf{R}}_\Delta ^{ - 1}{{{\bf{ R}}}^{ - 1}}} \right)^T}{{\bf{n}}_{{\rm{OAR}}}}{\Gamma _i}\left( {{{\bf{n}}_{{\rm{OAR}}}}} \right).
\end{align}

According to the generalized Lambertian emission   model \eqref{channel_gain}, the channel gain $g_i$ is bounded due to the bounded  $\Delta \alpha, \Delta \beta, \Delta \gamma$.
Specifically,  let  ${{{\underline g _{i}}}}\left( {{{\bf{n}}_{{\rm{OAR}}}}} \right)$ and  ${ {{\overline {{g}} }_i}}\left( {{{\bf{n}}_{{\rm{OAR}}}}} \right)$ denote the upper bound and lower bound of ${{{ g _{i}}}}\left( {{{\bf{n}}_{{\rm{OAR}}}}} \right)$, i.e.,
   \begin{subequations}
\begin{align}
&{\underline g _{i}}\left( {{{\bf{n}}_{{\rm{OAR}}}}} \right) \leq {g_{i}}\left( {{{\bf{n}}_{{\rm{OAR}}}}} \right) \leq {\overline g _{i}}\left( {{{\bf{n}}_{{\rm{OAR}}}}} \right),\\
&{{{\underline g _{i}}}}\left( {{{\bf{n}}_{{\rm{OAR}}}}} \right) = {{{\underline {\bf{q}} }_i^T}}{{\bf{n}}_{{\rm{OAR}}}}{\Gamma _i}\left( {{{\bf{n}}_{{\rm{OAR}}}}} \right),\\
 &{ {{\overline {{g}} }_i}}\left( {{{\bf{n}}_{{\rm{OAR}}}}} \right) = {\overline {\bf{q}} _i^T}{{\bf{n}}_{{\rm{OAR}}}}{\Gamma _i}\left( {{{\bf{n}}_{{\rm{OAR}}}}} \right),
 \end{align}
\end{subequations}
  where ${{{\underline {\bf{q}} }_i} } = {\lambda _i}{\left( {{{\bf{x}}_{{\rm{L}},i}} - {{\bf{r}}_{{\rm{UE}}}}} \right)}{\left( {{\bf{R}}_{ -  }^{ - 1}{{\bf{R}}^{ - 1}}} \right)}$,
${ {{\overline {\bf{q}} }_i}} = {\lambda _i}{\left( {{{\bf{x}}_{{\rm{L}},i}} - {{\bf{r}}_{{\rm{UE}}}}} \right)}{\left( {{\bf{R}}_{+ }^{ - 1}{{\bf{R}}^{ - 1}}} \right)}$, and ${{\bf{R}}_{ -  }}$ and ${{\bf{R}}_{ +  }}$ are the corresponding rotation  matrices to  achieve the minimum and maximum   of ${g_{i}}\left( {{{\bf{n}}_{{\rm{OAR}}}}} \right)$, respectively.
% Thus, the changing value of $g_i$ is bounded, i.e.,
%\begin{align}
% {\underline g _{i}}\left( {{{\bf{n}}_{{\rm{len}}}}} \right) \leq {g_{i}}\left( {{{\bf{n}}_{{\rm{len}}}}} \right) \leq {\overline g _{i}}\left( {{{\bf{n}}_{{\rm{len}}}}} \right),
%\end{align}
%where ${{{\underline g _{i}}}}\left( {{{\bf{n}}_{{\rm{len}}}}} \right) = {{{\underline {\bf{q}} }_i}}{{\bf{n}}_{{\rm{len}}}}{\Gamma _i}\left( {{{\bf{n}}_{{\rm{len}}}}} \right)$ and ${ {{\overline {\bf{q}} }_i}}\left( {{{\bf{n}}_{{\rm{len}}}}} \right) = {{{\overline{\bf q}}}^T}{{\bf{n}}_{{\rm{len}}}}{\Gamma _i}\left( {{{\bf{n}}_{{\rm{len}}}}} \right)$.
 Thus, the
channel gain ${g_{i}}\left( {{{\bf{n}}_{{\rm{OAR}}}}} \right)$ can be re-expressed as
\begin{align}
{g_{i}}\left( {{{\bf{n}}_{{\rm{OAR}}}}} \right) = {{\widehat g}_{i}}\left( {{{\bf{n}}_{{\rm{OAR}}}}} \right) + \Delta {g_{i}},
\end{align}
where ${{\widehat g}_{i}}\left( {{{\bf{n}}_{{\rm{OAR}}}}} \right) $ denotes the estimated CSI, i.e.,
\begin{align}
{\widehat g_{i}}\left( {{{\bf{n}}_{{\rm{OAR}}}}} \right)& = \frac{{ {{\overline g }_{i}\left( {{{\bf{n}}_{{\rm{OAR}}}}} \right)}+{{\underline g }_{i}\left( {{{\bf{n}}_{{\rm{OAR}}}}} \right)} }}{2} = \frac{{\left( {{{\underline {\bf{q}} }_i} + {{\overline {\bf{q}} }_i}} \right)}^T}{2}{{\bf{n}}_{{\rm{OAR}}}}{\Gamma _i}\left( {{{\bf{n}}_{{\rm{OAR}}}}} \right),
\end{align}
 and
 $\Delta {g_{i}}$ denotes the random CSI uncertainty, i.e., $\left| {\Delta {g_{i}}} \right| \leq \frac{{{{\overline g }_{i}\left( {{{\bf{n}}_{{\rm{OAR}}}}} \right)} - {{\underline g }_{i}}\left( {{{\bf{n}}_{{\rm{OAR}}}}} \right)}}{2}$.

Let  ${\bf{g}}\left( {{{\bf{n}}_{{\rm{OAR}}}}} \right) = {\left[ {{g_1}\left( {{{\bf{n}}_{{\rm{OAR}}}}} \right),...,{g_N}\left( {{{\bf{n}}_{{\rm{OAR}}}}} \right)} \right]^T}$ denote the channel gain
vector between the LEDs and UE, i.e.,
\begin{align}
{{\bf{g}}}\left( {{{\bf{n}}_{{\rm{OAR}}}}} \right) = { {\widehat {\bf{g}}}}\left( {{{{{\bf n}}}_{{\rm{OAR}}}}} \right) + \Delta {{\bf{g}}},
\end{align}
where ${\widehat {\bf{g}}}\left( {{{{{\bf n}}}_{{\rm{OAR}}}}} \right) = {\left[ {{{\widehat g}_1}\left( {{{{{\bf n}}}_{{\rm{OAR}}}}} \right),...,{{\widehat g}_N}\left( {{{{{\bf n}}}_{{\rm{OAR}}}}} \right)} \right]^T}$ is the estimated CSI vector, and $\Delta {{\bf{g}}} = {\left[ {\Delta {g_1},...,\Delta {g_N}} \right]^T}$ is the CSI uncertainty vector, which can be characterized by the following ellipsoidal region
\begin{align}
\Omega  \buildrel \Delta \over = \left\{ {\Delta {{\bf{g}}}|\Delta {\bf{g}}^T{{\bf{C}}}\Delta {{\bf{g}}} \le \upsilon} \right\},
\end{align}
where ${{\bf{C}}} = {\bf{C}}^T\succeq \mathbf{0}$ controls the extension of the  ellipsoid,  and
$\upsilon$ determines  the volume  of the ellipsoid. The parameters   ${{\bf{C}}}$ and $\upsilon$ are determined by the UE rotation   range.

Consequently, the received signal at the UE is given as
\begin{align}
{y_{{\rm{mob}}}} = {\left( {{{\widehat {\bf{g}}}\left( {{{{{\bf n}}}_{{\rm{OAR}}}}} \right)} + {\Delta {{\bf{g}}}}} \right)^T}{\bf{x}} + z.
\end{align}

Based on \eqref{rate}, the UE achievable rate   is given by
 \begin{align}\label{rate_B_A}
&{R_{{\rm{mob}}}}\left( {{{{{\bf n}}}_{{\rm{OAR}}}}} ,{\bf{p}} \right) = B{\log _2}\left( {1 + \frac{{{{\left| {{{\left( {{\widehat {\bf{g}}\left( {{{{{\bf n}}}_{{\rm{OAR}}}}} \right)} + {\Delta {\bf{g}}}} \right)}^T}{\bf{p}}} \right|}^2}{e^{1 + 2\left( {\alpha_0  + \gamma_0 \varepsilon_0} \right)}}}}{{2\pi B{\sigma ^2}}}} \right).
\end{align}

For the  random UE orientation,
the   total transmit power is minimized by jointly
optimizing the orientation vector ${{\bf{n}}_{{\rm{OAR}}}}$ and beamforming vector $\bf p$, which can be
formulated as
\begin{subequations}\label{robust_pro}
\begin{align}
 \mathop {\min }\limits_{{{{{{\bf n}}}_{{\rm{OAR}}}}} ,{\bf{p}}}  ~&\varepsilon {\left\| {\bf{p}} \right\|^2}\\
{\rm{s}}{\rm{.t}}{\rm{.}}~&{R_{{\rm{mob}}}}\left( {{{{{\bf n}}}_{{\rm{OAR}}}}} ,{\bf{p}} \right) \ge \bar R,\label{robust_pro_b}\\
~&{\left\| {{{{{\bf n}}}_{{\rm{OAR}}}}} \right\|^2} \le 1,\label{robust_pro_c}\\
~&{\left\| {\left( {{{\bf{e}}_1} + {{\bf{e}}_2}} \right) \odot {{{{\bf n}}}_{{\rm{OAR}}}}} \right\|^{\rm{2}}} \le {\sin ^2}\bar \theta ,\label{robust_pro_d}\\
~&\cos \bar \theta  \le {{{{\bf n}}}_{{\rm{OAR}}}^T}{{\bf{e}}_3} \le {\rm{1}},\label{robust_pro_e}\\
~&\Delta {{\bf{g}}} \in {\Omega },\label{robust_pro_f}\\
~&\sqrt {{P_i}} A \le {I_{{\rm{DC}}}},\forall i \in {\cal N}.\label{robust_pro_g}
\end{align}
\end{subequations}

Due to the fact that the orientation vector ${{\bf{n}}_{{\rm{OAR}}}}$ and beamforming vector $\bf p$ are coupled
together, problem \eqref{robust_pro} is non-convex. To deal with this issue, we exploit
the AO technique to optimize   ${{\bf{n}}_{{\rm{OAR}}}}$
and  $\bf p$  alternately.
Specifically, we  first decouple problem \eqref{robust_pro}
    into two subproblems, i.e., a robust beamforming subproblem and a PD  orientation optimization subproblem.

\subsection{Robust Beamforming Design Subproblem}
With a given  orientation vector ${{{{\bf n}}}_{{\rm{OAR}}}}$, i.e., given estimated channel gain $ {{\widehat {\bf{g}}}\left( {{{{{\bf n}}}_{{\rm{OAR}}}}} \right)}$, we first optimize a robust beamformer   to minimize the total transmit power. Specifically, the subproblem  is given as
\begin{subequations}\label{robust_pro2}
\begin{align}
 \mathop {\min }\limits_{{\bf{p}}}  ~&\varepsilon {\left\| {\bf{p}} \right\|^2}\\
{\rm{s}}{\rm{.t}}{\rm{.}}~&B{\log _2}\left( {1 + \frac{{{{\left| {{{\left( {{\widehat {\bf{g}}\left( {{{{{\bf n}}}_{{\rm{OAR}}}}} \right)} + {\Delta {\bf{g}}}} \right)}^T}{\bf{p}}} \right|}^2}{e^{1 + 2\left( {\alpha_0  + \gamma_0 \varepsilon_0 } \right)}}}}{{2\pi B{\sigma ^2}}}} \right) \ge \bar R,\label{robust_pro2_b}\\
~&\Delta {{\bf{g}}} \in {\Omega },\label{robust_pro2_c}\\
~&\sqrt {{P_i}} A \le {I_{{\rm{DC}}}},\forall i \in {\cal N}\label{robust_pro2_d}.
\end{align}
\end{subequations}

Moreover, by   defining ${{\bf{P}}} = {{\bf{p}}}{\bf{p}}^{\rm{T}} $, the constraint \eqref{robust_pro2_b} can be rewritten as
\begin{align}\label{S1_rate_1}
\Delta {\bf{g}}^T{\bf{P}}\Delta {{\bf{g}}} + 2\Delta {\bf{g}}^T{\bf{P}}{\widehat {\bf{g}}}\left( {{{{{\bf n}}}_{{\rm{OAR}}}}} \right) + {\widehat {\bf{g}}}\left( {{{{{\bf n}}}_{{\rm{OAR}}}}} \right)^T{\bf{P}}{\widehat {\bf{g}}}\left( {{{{{\bf n}}}_{{\rm{OAR}}}}} \right) \ge {c_1},
\end{align}
where ${c_1}=\left( {{2^{\frac{{\bar R}}{B}}} - 1} \right)\frac{{2\pi B{\sigma ^2}}}{{{e^{1 + 2\left( {\alpha_0  + \gamma_0 \varepsilon_0 } \right)}}}}$.

Due to the bounded constraint $\Delta {{\bf{g}}} \in {\Omega }$, the number of constraints in \eqref{S1_rate_1} is  infinite. Then,   based on the S-Procedure,  we conservatively transform the infinite constraints to  finite  linear matrix inequality constraints as detailed next.

%\textbf{Lemma} (S-Procedure): Let a function ${f_i}\left( {\bf{x}} \right),i = \left\{ {1,2} \right\}$, be defined as
%\begin{align}
%{f_i}\left( {\bf{x}} \right) = {{\bf{x}}^H}{{\bf{A}}_i}{\bf{x}} + 2{\mathop{\rm Re}\nolimits} \left\{ {{\bf{b}}_i^H{\bf{x}}} \right\} + {z_i},
%  \end{align}
%where ${\bf{x}} \in {{\mathbb{C}}^{N \times 1}},{{\bf{A}}_i} \in {{\mathbb{C}}^{N \times N}},{{\bf{A}}_i} = {\bf{A}}_i^H,{{\bf{b}}_i} \in {{\mathbb{C}}^{N \times 1}},{z_i} \in {\mathbb{R}}$. Then,  ${f_1}\left( {\bf{x}} \right) \le 0 \Rightarrow {f_2}\left( {\bf{x}} \right) \le 0$ holds if and only if there exits a variable $\eta   \ge 0$ such that
%\begin{align}\label{S1_rate_2}
%\left[ {\begin{array}{*{20}{c}}
%{{{\bf{A}}_1}}&{{{\bf{b}}_1}}\\
%{{\bf{b}}_1^H}&{{z_1}}
%\end{array}} \right] + \eta \left[ {\begin{array}{*{20}{c}}
%{{{\bf{A}}_2}}&{{{\bf{b}}_2}}\\
%{{\bf{b}}_2^H}&{{z_2}}
%\end{array}} \right] \succeq 0,
%  \end{align}
%where \eqref{S1_rate_2} assumes that there exists a point $\widehat {\bf{x}}$ such that ${f_k}\left( {\widehat {\bf{x}}} \right) \le 0$.

By using the S-Procedure {\cite{Boyd_Cambridge_2004}}, the   constraints in \eqref{S1_rate_1} can be converted  as
\begin{align}\label{SS1_rate_1}
\left[ {\begin{array}{*{20}{c}}
{{\bf{P}} + \eta {\bf{C}}}&{{\bf{P}}\widehat {\bf{g}}\left( {{{\bf{n}}_{{\rm{OAR}}}}} \right)}\\
{\widehat {\bf{g}}{{\left( {{{\bf{n}}_{{\rm{OAR}}}}} \right)}^T}{\bf{P}}}&\begin{array}{c}
\widehat {\bf{g}}{\left( {{{\bf{n}}_{{\rm{OAR}}}}} \right)^T}{\bf{P}}\widehat {\bf{g}}\left( {{{\bf{n}}_{{\rm{OAR}}}}} \right)\\
 - {c_1} - \eta v
\end{array}
\end{array}} \right] \succeq 0.
  \end{align}

Furthermore, by ignoring the rank-one constraint of $\bf P$ due to SDR, we obtain the following conservative approximation  problem as follows.
\begin{subequations}\label{min_power_4}
\begin{align}
 \mathop {\min }\limits_{{\bf{P}},c_2}  &~\varepsilon{\rm{Tr}}\left( {\bf{P}} \right)\\
 {\rm{s.t.}}&~ \left[ {\begin{array}{*{20}{c}}
{{\bf{P}} + \eta {\bf{C}}}&{{\bf{P}}\widehat {\bf{g}}\left( {{{\bf{n}}_{{\rm{OAR}}}}} \right)}\\
{\widehat {\bf{g}}{{\left( {{{\bf{n}}_{{\rm{OAR}}}}} \right)}^T}{\bf{P}}}&\begin{array}{c}
\widehat {\bf{g}}{\left( {{{\bf{n}}_{{\rm{OAR}}}}} \right)^T}{\bf{P}}\widehat {\bf{g}}\left( {{{\bf{n}}_{{\rm{OAR}}}}} \right)\\
 - {c_1} - \eta v
\end{array}
\end{array}} \right] \succeq 0, \label{min_power_4b}\\
 &~{\rm{Tr}}\left( {{\bf{P}}{{\bf{e}}_i}{\bf{e}}_i^T} \right) \le \frac{{I_{{\rm{DC}}}^2}}{A},\forall i \in {\cal N},\label{min_power_4c}\\
 &~{\bf{P}}\succeq 0.\label{min_power_4d}
% &~{\left\| {{{\bf{n}}_{{\rm{len}}}}} \right\|^2} \le 1,\\
%&~{{\bf{n}}_{{\rm{len}}}}^T\Lambda {{\bf{n}}_{{\rm{len}}}} \le {\sin ^2}\bar \theta ,\\
%&~\cos \bar \theta  \le {{{{\bf{n}}_{{\rm{len}}}}}^T}{\bf{c}} \le {\rm{1}}.
\end{align}
\end{subequations}

Therefore, problem  \eqref{min_power_4} can be solved by the interior-point algorithm \cite{Ye_1997}\cite{Grant_2009}. It  can be calculated that the complexity of problem  \eqref{min_power_4} is ${\mathcal O}\left( {{{\left( {N + 2} \right)}^4}{{\left( {N + 1} \right)}^{{1 \mathord{\left/
 {\vphantom {1 2}} \right.
 \kern-\nulldelimiterspace} 2}}}\log \left( {{1 \mathord{\left/
 {\vphantom {1 {\zeta}_3 }} \right.
 \kern-\nulldelimiterspace} {\zeta}_3 }} \right)} \right)$, where ${\zeta}_3 > 0$ is the solution accuracy {\cite{Luo_ISPM_2010}}. When the  rank of the optimal solution is one, we can compute the optimal beamforming vectors by eigenvalue decomposition. Otherwise, we can use the Gaussian randomization procedure to generate a high-quality feasible beamformer vector \cite{Luo_ISPM_2010}.

 \subsection{PD Orientation Optimization Subproblem}

 In this subsection, we will optimize the orientation vector
${{\bf{n}}_{{\rm{OAR}}}}$ when the {transmit} beamforming vector ${{\bf{p}}}$ is fixed.
  Specifically,  the  orientation vector
${{\bf{n}}_{{\rm{OAR}}}}$   optimization subproblem can be formulated as follows.
\begin{subequations}\label{robust_pro1_o}
\begin{align}
 \mathop {\min }\limits_{{{{{{\bf n}}}_{{\rm{OAR}}}}}}  ~&\varepsilon {\left\| {\bf{p}} \right\|^2}\label{robust_pro1_o_a}\\
{\rm{s}}{\rm{.t}}{\rm{.}}~&B{\log _2}\left( {1 + \frac{{{{\left| {{{\left( {{\widehat {\bf{g}}\left( {{{{{\bf n}}}_{{\rm{OAR}}}}} \right)} } \right)}^T}{\bf{p}}} \right|}^2}{e^{1 + 2\left( {\alpha_0  + \gamma_0 \varepsilon_0 } \right)}}}}{{2\pi B{\sigma ^2}}}} \right) \ge \bar R,\label{robust_pro1_o_b}\\
~&{\left\| {{{{{\bf n}}}_{{\rm{OAR}}}}} \right\|^2} \le 1,\label{robust_pro1_o_c}\\
~&{\left\| {\left( {{{\bf{e}}_1} + {{\bf{e}}_2}} \right) \odot {{{{\bf n}}}_{{\rm{OAR}}}}} \right\|^{\rm{2}}} \le {\sin ^2}\bar \theta ,\label{robust_pro1_o_d}\\
~&\cos \bar \theta  \le {{{{\bf n}}}_{{\rm{OAR}}}^T}{{\bf{e}}_3} \le {\rm{1}}.\label{robust_pro1_o_e}
%~&\Delta {{\bf{g}}} \in {\Omega },\label{robust_pro1_o_f}
\end{align}
\end{subequations}

With the fixed ${\bf{p}}$, the objective function \eqref{robust_pro1_o_a} does not depend on ${{{\bf n}}_{{\rm{OAR}}}}$. However, when the orientation vector ${{{\bf n}}_{{\rm{OAR}}}}$ can achieve the   maximum ${{{\left( {\widehat {\bf{g}}\left( {{{\bf{n}}_{{\rm{OAR}}}}} \right)} \right)}^T}{\bf{p}}}$, the power of  beamforming vector ${\bf{p}}$ is the minimum. Thus, the orientation optimization subproblem can be reformulated as
\begin{subequations}\label{robust_pro1}
\begin{align}
 \mathop {\max }\limits_{{{{{\bf n}}_{{\rm{OAR}}}}}}  &~{{{\left( {\widehat {\bf{g}}\left( {{{{\bf n}}_{{\rm{OAR}}}}} \right) } \right)}^T}{\bf{p}}} \label{robust_pro1_a}\\
{\rm{s}}{\rm{.t}}{\rm{.}}&~{\left( {\widehat {\bf{g}}\left( {{{\bf{n}}_{{\rm{OAR}}}}} \right)} \right)^T}{\bf{p}} \ge {c_2},\label{robust_pro1_b}\\
&\eqref{robust_pro1_o_c},\eqref{robust_pro1_o_d},\eqref{robust_pro1_o_e}\nonumber.
%&~\Delta {{\bf{g}}} \in {\Omega }.
\end{align}
\end{subequations}

By substituting ${\widehat g_{i}}\left( {{{\bf{n}}_{{\rm{OAR}}}}} \right) = \frac{{\left( {{{\underline {\bf{q}} }_i} + {{\overline {\bf{q}} }_i}} \right)}}{2}{{\bf{n}}_{{\rm{OAR}}}}{\Gamma _i}\left( {{{\bf{n}}_{{\rm{OAR}}}}} \right)$ into ${{{\left( {\widehat {\bf{g}}\left( {{{\bf{n}}_{{\rm{OAR}}}}} \right)} \right)}^T}{\bf{p}}}$,   constraint \eqref{robust_pro1_b} can be reformulated as
\begin{align}
\sum\limits_{n = 1}^N {\sqrt {{p_n}} \frac{{{{\left( {{{\underline {\bf{q}} }_n} + {{\overline {\bf{q}} }_n}} \right)}^T}}}{2}{{\bf{n}}_{{\rm{OAR}}}}{\Gamma _n}\left( {{{\bf{n}}_{{\rm{OAR}}}}} \right)} \ge {c_2},
\end{align}
where ${c_2} = \sqrt {\left( {{2^{\frac{{\bar R}}{B}}} - 1} \right)\frac{{2\pi B{\sigma ^2}}}{{{e^{1 + 2\left( {\alpha_0  + \gamma_0 \varepsilon_0 } \right)}}}}} $.

Thus,  the objective function \eqref{robust_pro1_a} and constraint  \eqref{robust_pro1_b} with orientation vector ${{{{\bf n}}}_{{\rm{OAR}}}}$ can  be reformulated as
\begin{subequations}\label{robust_pro3}
\begin{align}
 \mathop {\max }\limits_{{{{{\bf n}}_{{\rm{OAR}}}}}}  &~{\sum\limits_{n = 1}^N {\sqrt {{p_n}} \frac{{{{\left( {{{\underline {\bf{q}} }_n} + {{\overline {\bf{q}} }_n}} \right)}^T}}}{2}{{\bf{n}}_{{\rm{OAR}}}}{\Gamma _n}\left( {{{\bf{n}}_{{\rm{OAR}}}}} \right)}}  \label{robust_pro3_a}\\
{\rm{s}}{\rm{.t}}{\rm{.}}&~\sum\limits_{n = 1}^N {\sqrt {{p_n}} \frac{{{{\left( {{{\underline {\bf{q}} }_n} + {{\overline {\bf{q}} }_n}} \right)}^T}}}{2}{{\bf{n}}_{{\rm{OAR}}}}{\Gamma _n}\left( {{{\bf{n}}_{{\rm{OAR}}}}} \right)} \ge {c_2},\label{robust_pro3_b}\\
&~\eqref{robust_pro1_o_c},
\eqref{robust_pro1_o_d},\eqref{robust_pro1_o_e}\nonumber.
%&~\Delta {{\bf{g}}} \in {\Omega }.
\end{align}
\end{subequations}

Due to the integer and indicator function ${\Gamma _n}\left( {{{\bf{n}}_{{\rm{OAR}}}}} \right)$ in both objective function \eqref{robust_pro3_a} and constraint \eqref{robust_pro3_b},  problem \eqref{robust_pro3} is a non-linear non-smooth problem, which is
hard to address \cite{Luenberger_1984}.
To overcome the  challenging issue, we propose an  alternating optimization and projection method to  solve  problem \eqref{robust_pro3} iteratively.
To be more specific, during   the $k$th iteration,  we relax      problem \eqref{robust_pro3}   as
\begin{subequations}\label{robust_prob1_2}
\begin{align}
 \mathop {\max }\limits_{{{{{\bf n}}_{{\rm{OAR}}}}}}  &~\sum\limits_{j \in {\cal M}} {\sqrt {{p_n}} \frac{{{{\left( {{{\underline {\bf{q}} }_n} + {{\overline {\bf{q}} }_n}} \right)}^T}}}{2}{{\bf{n}}_{{\rm{OAR}}}}}   \label{robust_pro1_2_a}\\
{\rm{s}}{\rm{.t}}{\rm{.}}&~\sum\limits_{j \in {\cal M}} {\sqrt {{p_n}} \frac{{{{\left( {{{\underline {\bf{q}} }_n} + {{\overline {\bf{q}} }_n}} \right)}^T}}}{2}{{\bf{n}}_{{\rm{OAR}}}}}  \ge {c_2},\\
&~\eqref{robust_pro1_o_c},
\eqref{robust_pro1_o_d},\eqref{robust_pro1_o_e},\nonumber
\end{align}
\end{subequations}
where ${\mathcal M} \buildrel \Delta \over = \left\{ {\forall j \in {\mathcal N}\left| {{\Gamma _j}\left( {\widetilde {\bf{n}}_{{\rm{OAR}}}^{\left[ k \right]}} \right)  \ne 0} \right.} \right\}$ denotes the LEDs set. Then,  by applying the standard interior-point algorithm \cite{Ye_1997},
  problem \eqref{robust_prob1_2} can be efficiently
solved and the solution $\widetilde {\bf{n}}_{{\rm{OAR}}}^{\left[ k \right]}$  can be obtained.
 Then, the LED set $\cal M$ is updated.
Note that the value of the indicator function ${\Gamma _i}\left( {{{\widetilde {\bf{n}}_{{\rm{OAR}}}^{\left[ 0 \right]}}}} \right)$  is derived by following the principle:  If $\frac{{{\bf{d}}_i^T{\widetilde {\bf{n}}_{{\rm{OAR}}}^{\left[ 0 \right]}}}}{{{d_i}}} < \cos \left( {{\psi _{{\rm{FOV}}}}} \right)$, ${\Gamma _i}\left( {{{\widetilde {\bf{n}}_{{\rm{OAR}}}^{\left[ 0 \right]}}}} \right)= 0$; otherwise, ${\Gamma _i}\left( {{{\widetilde {\bf{n}}_{{\rm{OAR}}}^{\left[ 0 \right]}}}} \right)= 1$.

\subsection{AO algorithm for Random UE orientation}

The proposed AO algorithm for jointly robust beamforming design and orientation optimization problem
 is summarized in Algorithm 3.
We first initialize a given receiving orientation vector ${\bf{n}}_{{\rm{OAR}}}^{\left[ 0 \right]}$, and obtain a robust beamforming vector
${\bf{P}}^{\left[ 1 \right]}$ by solving subproblem   \eqref{min_power_4}. Then,
  we update ${\bf{n}}_{{\rm{OAR}}}^{\left[ 1 \right]}$ by solving subproblem \eqref{robust_prob1_2}. By iteratively calculating ${\bf{n}}_{{\rm{OAR}}}^{\left[ k \right]}$ and ${\bf{P}}^{\left[ k \right]}$, the proposed Algorithm 3 proceeds until convergence.
  The details of algorithm are given as follows.
\begin{algorithm}[H]
    \caption{AO Algorithm for Problem \eqref{robust_pro}}
    \label{AO_robust}
    \begin{algorithmic}[1]
%        \Require ${{\bf{P}}^{[0]}}$
%        \Ensure ${{\bf{P}}^ * },{{\bf{n}}_{{\rm{len}}}^ * }$
%        \State \textbf{Given} the convergence criterion ${\kappa }  > 0$, set angel error threshold
%         $\bar \alpha, \bar \beta, \bar \gamma$;
        \State \textbf{Initialize:} given the convergence criterion ${\kappa }  > 0$, choose an initial orientation vector ${\bf{n}}_{{\rm{OAR}}}^{\left[ 0 \right]}$, set the  random angle errors
         $\Delta \alpha, \Delta \beta, \Delta \gamma$, and set iteration number   $ k = 1$;
        \State Calculating ${\bf{P}}^{\left[ k \right]}$ based on \eqref{min_power_4};
        \State Calculating ${\bf{n}}_{{\rm{OAR}}}^{\left[ k \right]}$ based on \eqref{robust_prob1_2};
         \While{$\left\| {{\bf{n}}_{{\rm{OAR}}}^{\left[ k \right]} - {\bf{n}}_{{\rm{OAR}}}^{\left[ k-1 \right]}} \right\| \ge \kappa $}
         \State $ k \leftarrow k+1 $;
         \State Go back to Step 2;
          \EndWhile
         \State Obtain ${{\bf{P}}^ * } \leftarrow {{\bf{P}}^{[k]}},{\bf{n}}_{{\rm{OAR}}}^ *  \leftarrow {\bf{n}}_{{\rm{OAR}}}^{\left[ k \right]}$.
    \end{algorithmic}
\end{algorithm}

 This algorithm includes two main methods solving a robust beamforming design subproblem in Step 2 and a PD orientation optimization subproblem in Step 3.
To obtain robust beamforming vector ${\bf{P}}^{\left[ 1 \right]}$ and PD orientation vector ${\bf{n}}_{{\rm{OAR}}}^{\left[ 0 \right]}$, we use the interior-point algorithm.  In conclusion, the total computational complexity of the proposed AO algorithm is approximately  ${\cal O}\left( {\max {{\left\{ {N + 2,4} \right\}}^4}{\left({N+3}\right)^{{1 \mathord{\left/
 {\vphantom {1 2}} \right.
 \kern-\nulldelimiterspace} 2}}}\log \left( {{1 \mathord{\left/
 {\vphantom {1 {{\zeta _4}}}} \right.
 \kern-\nulldelimiterspace} {{\zeta _4}}}} \right)} \right)$, where ${\zeta _4} > 0$ is the solution accuracy {\cite{Luo_ISPM_2010}}.

\section{Simulation Results and Discussions}
In this section, simulation results are presented to demonstrate the effectiveness of our  proposed BO optimization strategies. Consider a  VLC system in a room with size of $\left(6 \times 6 \times 3\right)~\rm{m^3}$,
and one corner of the room is modeled as  the origin $(0, 0, 0)$ of the Cartesian
coordinate system ($X, Y, Z$). The person moves from $(0,0)$ to $(6, 6)$ along the diagonal line of the room. The VLC transmitter  contains $9$ LEDs, and  locations of LEDs are listed in Fig. 5. Moreover, the basic parameters of the VLC system are listed in Table II. Note that in the case of non-OAR, we only
solve the beamforming design problem without optimizing the PD orientation vector ${{\bf{n}}_{{\rm{OAR}}}}$, { i.e., the PD orientation ${{\bf{n}}_{{\rm{OAR}}}}$ is fixed as ${\left[ {0,0,1} \right]^T}$ in the UE coordinate system}.
\begin{figure}[H]
      \centering
	\includegraphics[width=7.2cm]{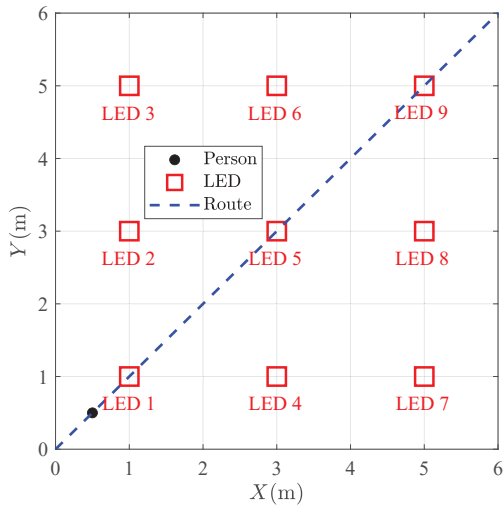}
      \vskip-0.2cm\centering
 \caption{ The schematic of LEDs and User location. }
  \label{The schematic of LEDs and User location} %% label for entire figure
\end{figure}

	\begin{table}[H]
	\centering
	\caption{\normalsize Basic parameters }
	\begin{threeparttable}
		\begin{tabular}{|l|l|}
			\hline
			%\multicolumn{2}{|c|}{VLC Link}\\
			%\hline
			Definition  & Value   \\
			\hline
			Room dimensions $(W \times L \times H)$  & {$\left(6 \times 6 \times 3\right)~\rm{m^3}$} \\
			\hline
			FoV~${\psi _{{\rm{FOV}}}}$ & ${60 ^ \circ }$  \\
			\hline
			Detector area of PD~ $A_{\rm{PD}}$ & {$1 ~{\rm{c}}{{\rm{m}}^2}$}  \\
			\hline
			Bandwidth ~$B$ & {per unit bandwidth\tnote{*} }    \\
			\hline
			Average electrical noise power ~${\sigma ^2}$ & {$-98.82  ~\rm{dBm}$}   \\
			\hline
			%Lens refractive index $n$         & $1.5$        \\
			%\hline
			Half power angle ~${\phi _{{1 \mathord{\left/
							{\vphantom {1 2}} \right.
							\kern-\nulldelimiterspace} 2}}}$&${60 ^ \circ }$ \\
			\hline
			\tabincell{l} {DC biasing  $I_{\rm{DC}}$ }&{$1$ A}  \\
			\hline
		\end{tabular}
		
		\begin{tablenotes}
			\footnotesize
			\item[*] Generally, the   bandwidth of VLC   system  ranges from $10~{\rm MHZ}$ to $600~{\rm MHZ}$ \cite{Pathak_ICST_2015},\cite{Ma_ITWC_2020}.
		\end{tablenotes}
	\end{threeparttable}
	
\end{table}

\subsection{Performance of OAR  System of Fixed  UE Orientation
}

\begin{figure}[!htp]
      \centering
	\includegraphics[width=7.4cm]{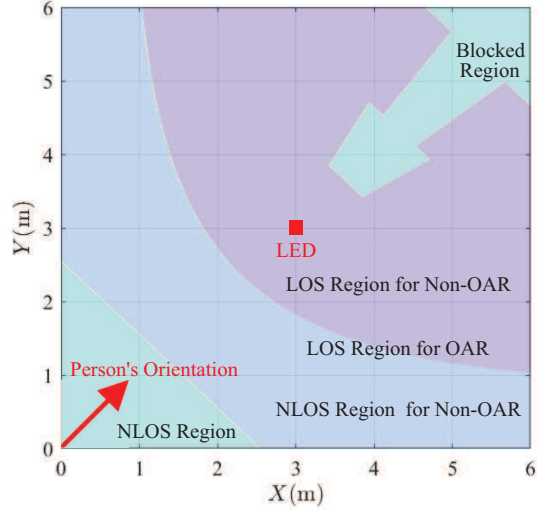}
      \vskip-0.2cm\centering
 \caption{ The schematic of the LOS, NLOS and blocked regions. }
  \label{BOLCK_graph} %% label for entire figure
\end{figure}

Fig.~\ref{BOLCK_graph} depicts the LOS, NLOS and blocked regions of the proposed VLC  system. In this scenario, a single LED is placed at the $(3,3)$ coordinate of the room, and  the user's orientation is depicted in Fig.~\ref{BOLCK_graph}.
We observe that the green area at the top right corner shows the region blocked by human body. The purple area depicts the LOS region for non-OAR UE. The  LOS region for the OAR UE is jointly described by the purple and blue areas. The blue area depicts the NLOS region of the UE equipped with non-OAR. Moreover, the green area at the left bottom shows the NLOS region for both OAR and non-OAR UE.

\begin{figure}[!htp]
      \centering
	\includegraphics[width=7.5cm]{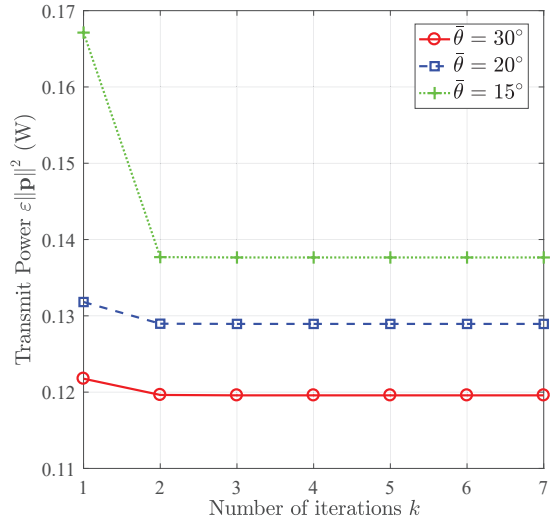}
      \vskip-0.2cm\centering
 \caption{ Transmit power $\varepsilon {\left\| {\bf{p}} \right\|^2}$ versus the number of iterations $k$, with minimum rate threshold ${\bar R} = 5 \left( { \rm bits/sec/Hz} \right)$. }
  \label{converge_graph} %% label for entire figure
\end{figure}

Fig.~\ref{converge_graph} plots the transmit power $\varepsilon {\left\| {\bf{p}} \right\|^2}$ versus the number of iterations $k$ for Algorithm 2  with minimum rate threshold ${\bar R} = 5 \left( { \rm bits/sec/Hz} \right)$. It can be observed that the proposed Algorithm 2 converges within three or four iterations,  which demonstrates its effectiveness. Moreover, the total transmit power $\varepsilon {\left\| {\bf{p}} \right\|^2}$  decreases with the increase  of the elevation angle  threshold $\bar \theta$. This is due to the fact that the elevation angle  with higher threshold has more degrees of freedom and can better capture the transmit light signal. Thus, less transmit power is required when the UE is equipped with OAR.

% while ${\left\| {\bf{p}} \right\|^2}$ of theta threshold $\bar \theta=30^\circ$ is lower than that of  elevation angle threshold $\bar \theta=15^\circ$. This is because when the elevation angle threshold is higher, less transmit power needed for user equipped with OAR.

\begin{figure}[!htp]
%\centering
%\subfigure[]{
%\centering
%\begin{minipage}[b]{0.5\textwidth}
%\includegraphics[width=7.5cm]{Figures/led9_threshold.eps}
%\end{minipage}
%}
%\subfigure[]{
%\centering
%\begin{minipage}[b]{0.5\textwidth}
%\includegraphics[width=7.5cm]{Figures/led9_ratethred.eps}
%\end{minipage}
%}
    \begin{minipage}[b]{0.45\textwidth}
      \centering
\includegraphics[height=7.5cm,width=7.5cm]{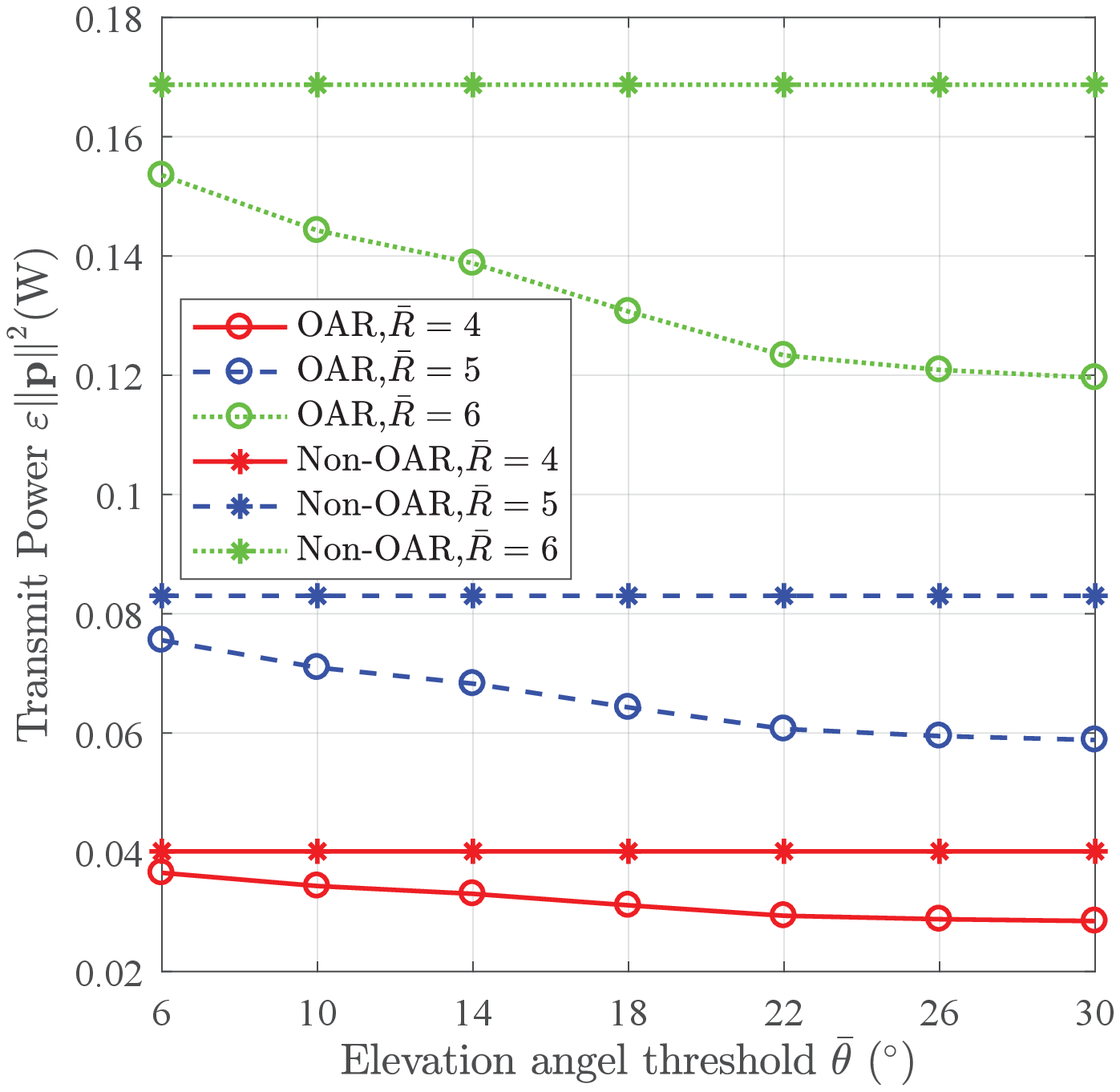}
      \vskip-0.2cm\centering {\footnotesize (a)}
    \end{minipage}\hfill
    \begin{minipage}[b]{0.45\textwidth}
      \centering
\includegraphics[height=7.5cm,width=7.5cm]{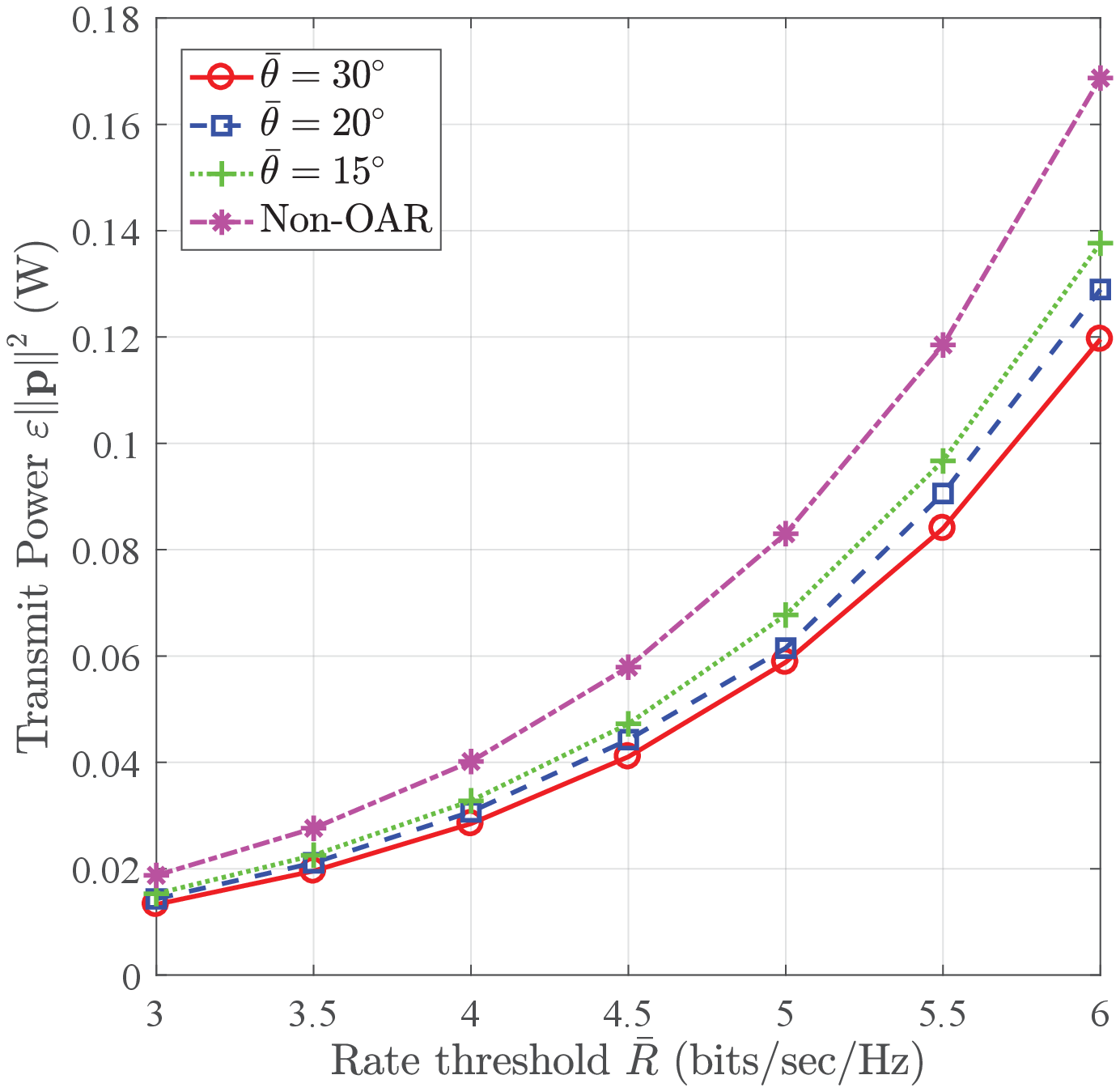}
      \vskip-0.2cm\centering {\footnotesize (b)}
    \end{minipage}\hfill
 \caption{(a)~Transmit power $\varepsilon {\left\| {\bf{p}} \right\|^2}$ versus the theta threshold with different rate threshold ${\bar R}$;
 (b)~Transmit power $\varepsilon {\left\| {\bf{p}} \right\|^2}$  versus rate threshold with different elevation angle threshold
 $\bar \theta$.}
  \label{threshold_graph} %% label for entire figure
\end{figure}

Fig.~\ref{threshold_graph} (a) illustrates the transmit power $\varepsilon {\left\| {\bf{p}} \right\|^2}$ versus elevation angle threshold with different minimum rate requirements ${\bar R} = 4 \left( { \rm bits/sec/Hz} \right)$, $5 \left( { \rm bits/sec/Hz} \right)$ and $ 6 \left( { \rm bits/sec/Hz} \right)$. It can be observed that  the transmit power $\varepsilon {\left\| {\bf{p}} \right\|^2}$ decreases as the elevation angle threshold increases,  for the OAR case, which is  same as the result shown in Fig. \ref{converge_graph}. For the case  without OAR, the transmit power is constant as the elevation angle threshold ${\bar \theta}$ increases. Moreover, comparing with the non-OAR scheme, the total transmit power $\varepsilon {\left\| {\bf{p}} \right\|^2}$ with OAR is lower. This is because  OAR can provide  more channel gain.
Fig.~\ref{threshold_graph} (b) plots the transmit power $\varepsilon {\left\| {\bf{p}} \right\|^2}$ versus minimum rate requirement ${\bar R} $  with different elevation angle thresholds $\bar \theta=15^\circ$, $20^\circ$ and $30^\circ$. We observe that  the transmit power $\varepsilon{\left\| {\bf{p}} \right\|^2}$ increases with rate threshold
${\bar R}$. In addition, it can be observed that for the case of OAR, transmit power  $\varepsilon{\left\| {\bf{p}} \right\|^2}$ of the threshold $\bar \theta=30^\circ$ is lower than  the case of $\bar \theta=20^\circ$, which is, in turn,  lower than that  for  the case of $\bar \theta=15^\circ$. Obviously,  the non-OAR scheme consumes more power than the OAR case  under the same rate constraint.

%\begin{figure}[!htp]
%\centering
%\subfigure[]{
%\begin{minipage}[b]{0.5\textwidth}
%\includegraphics[width=7.5cm]{Figures/led9_X_rate_1230.eps}
%\end{minipage}
%}
% \caption{(a)~~Achievable rate ${R}$  versus the $X$ coordinate of user's location with minimum rate threshold ${\bar R} = 5 \left( { \rm bits/sec/Hz} \right)$.}
%  \label{location_graph} %% label for entire figure
%\end{figure}
%
%\begin{figure}[!htp]
%\centering
%\ContinuedFloat
%\subfigure[]{
%\begin{minipage}[b]{0.5\textwidth}
%\includegraphics[width=7.6cm]{Figures/led9_X_power_1230.eps}
%\end{minipage}
%}
% \caption{(b)~Transmit power ${\left\| {\bf{p}} \right\|^2}$   versus the $X$ coordinate of user's location with minimum rate threshold ${\bar R} = 5 \left( { \rm bits/sec/Hz} \right)$.}
%  \label{location_b_graph} %% label for entire figure
%\end{figure}

\begin{figure}[!htp]
%\centering
%\subfigure[]{
%\begin{minipage}[b]{0.5\textwidth}
%\includegraphics[width=7.5cm]{Figures/led9_X_rate_1230.eps}
%\end{minipage}
%}
%\subfigure[]{
%\begin{minipage}[b]{0.5\textwidth}
%\includegraphics[width=7.6cm]{Figures/led9_X_power_1230.eps}
%\end{minipage}
%}
    \begin{minipage}[b]{0.45\textwidth}
      \centering
\includegraphics[height=7.5cm,width=7.5cm]{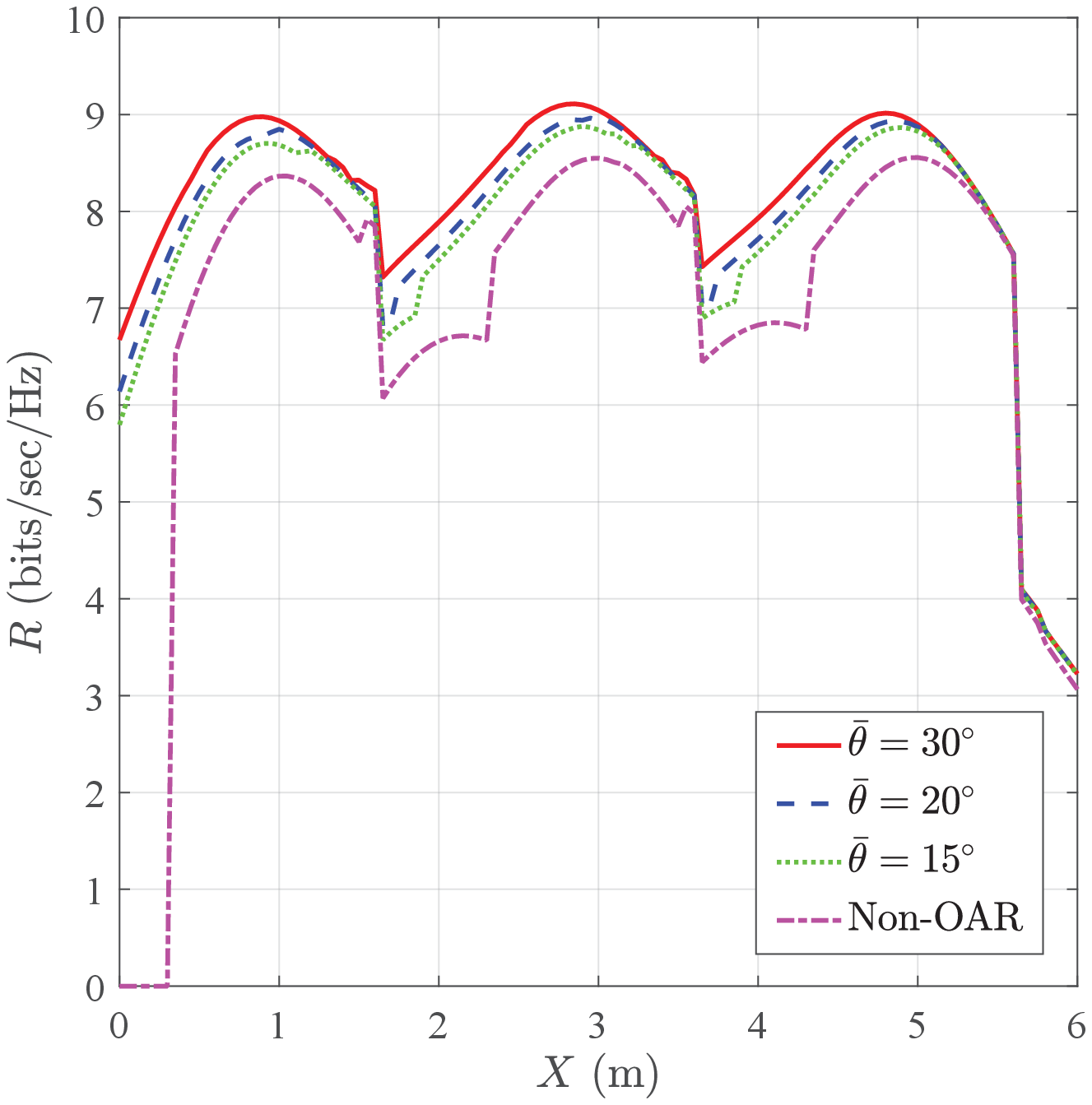}
      \vskip-0.2cm\centering {\footnotesize (a)}
    \end{minipage}\hfill
    \begin{minipage}[b]{0.45\textwidth}
      \centering
\includegraphics[height=7.5cm,width=7.5cm]{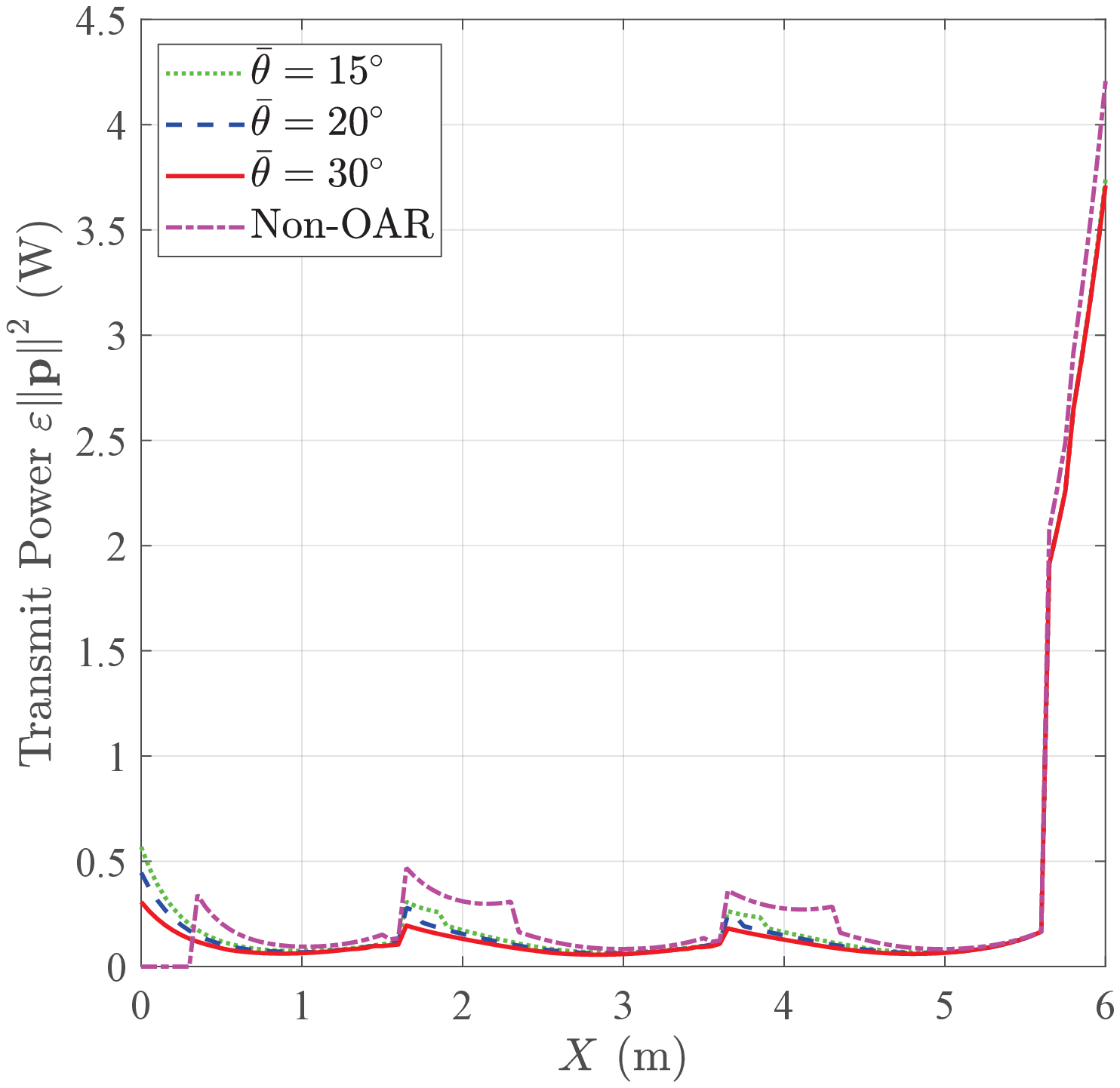}
      \vskip-0.2cm\centering {\footnotesize (b)}
    \end{minipage}\hfill
 \caption{(a)~~Achievable rate ${R}$  versus the $X$ coordinate of user's location; (b)~Transmit power $\varepsilon {\left\| {\bf{p}} \right\|^2}$   versus the $X$ coordinate of user's location with minimum rate threshold ${\bar R} = 1 \left( { \rm bits/sec/Hz} \right)$.}
  \label{location_graph} %% label for entire figure
\end{figure}

Fig.~\ref{location_graph}  (a) and (b) show the achievable rate ${R}$ and transmit power $\varepsilon {\left\| {\bf{p}} \right\|^2}$ versus the $X$ coordinate of the user's location, under the minimum rate requirement ${\bar R} = 1 \left( { \rm bits/sec/Hz} \right)$.
In Fig.~\ref{location_graph}  (a), it can be observed that the achievable rate roughly changes periodically with three high points at {$X=1~{\rm{m}}$, $X=3~{\rm{m}}$ and $X=5~{\rm{m}}$} as the $X$ axis of the user changes from 0 to 6. Meanwhile, the achievable rate ${R}$ drops off sharply three times at {$X=1.65~{\rm{m}}$, $X=3.65~{\rm{m}}$ and $X=5.65~{\rm{m}}$}. This is because the  signal suffers from human blockage as the user moves away from LED1, LED5 and LED9. We can observe that the achievable rate ${R}$   for the case of  elevation angle threshold $\bar \theta=30^\circ$ is higher than that of $\bar \theta=20^\circ$, which is, in turn,  higher than that of $\bar \theta=15^\circ$. Furthermore, the achievable rate $R$ of OAR with three different elevation angle thresholds is higher than that of the non-OAR scheme.
Fig.~\ref{location_graph}  (b) shows  that the transmit power has  roughly periodic behavior with three low points at {$X=1~{\rm{m}}$, $X=3~{\rm{m}}$ and $X=5~{\rm{m}}$} as the $X$ coordinate of the user changes from 0 to 6. Meanwhile, the transmit power $\varepsilon {\left\| {\bf{p}} \right\|^2}$ increases sharply three times at {$X=1.65~{\rm{m}}$, $X=3.65~{\rm{m}}$ and $X=5.65{\rm{m}}$}. This is again due to signal suffering from  human blockage. Other conclusions from this figure are the same as those of  Fig.~\ref{location_graph} (a).

\subsection{Performance of OAR system for Random UE Orientation}
In the following, we compare the proposed robust joint BO design scheme with the non-robust  joint BO design scheme, which does not handle random UE orientation, and the perfect joint BO design scheme, where  the random   angle errors $\Delta \alpha$, $\Delta \beta$ and $\Delta \gamma$ are zero.

\begin{figure}[!htp]
      \centering
	\includegraphics[width=7.5cm]{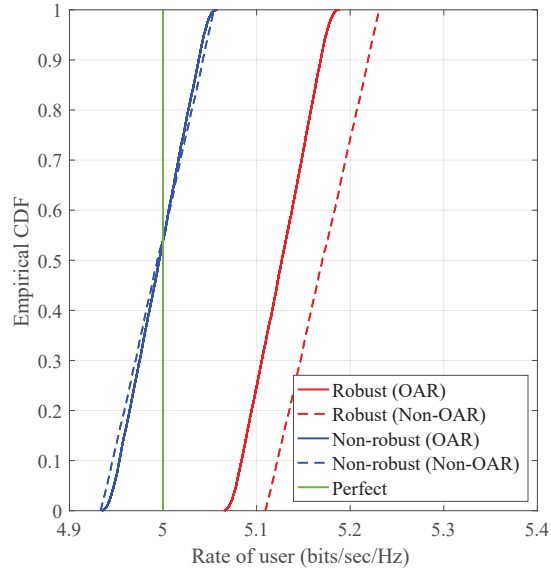}
      \vskip-0.2cm\centering
 \caption{ The empirical CDF of rate, with elevation angle threshold $\bar \theta=30^\circ$, minimum rate threshold ${\bar R} = 5 \left( { \rm bits/sec/Hz} \right)$ and random angle errors $\Delta \alpha  \in \left[ { - {{0.5}^ \circ },{{0.5}^ \circ }} \right)$, $ \Delta \beta  \in \left[ { - {{0.5}^ \circ },{{0.5}^ \circ }} \right)$, $ \Delta \gamma  \in \left[ { - {{0.5}^ \circ },{{0.5}^ \circ }} \right)$. }
  \label{CDF_graph} %% label for entire figure
\end{figure}

{Fig.~\ref{CDF_graph} shows the cumulative distribution function (CDF) of the achievable rate, where the elevation angle threshold $\bar \theta=30^\circ$, the minimum rate threshold ${\bar R} = 5 \left( { \rm bits/sec/Hz} \right)$ and random angle errors $\Delta \alpha  \in \left[ { - {{0.5}^ \circ },{{0.5}^ \circ }} \right)$, $ \Delta \beta  \in \left[ { - {{0.5}^ \circ },{{0.5}^ \circ }} \right)$, $ \Delta \gamma  \in \left[ { - {{0.5}^ \circ },{{0.5}^ \circ }} \right)$, respectively.  The CDF of rate is over 10000 random channel realizations  by Monte-Carlo simulation.} On one hand,  both the robust design with OAR and non-OAR satisfy the minimum rate requirement. On the other hand, the non-robust with OAR and non-OAR schemes cannot always guarantee the minimum rate constraint. It can also be verified that the proposed robust design with OAR is less conservative than the robust design with non-OAR.

\begin{figure}[!htp]
    \begin{minipage}[b]{0.45\textwidth}
      \centering
\includegraphics[height=7.5cm,width=7.5cm]{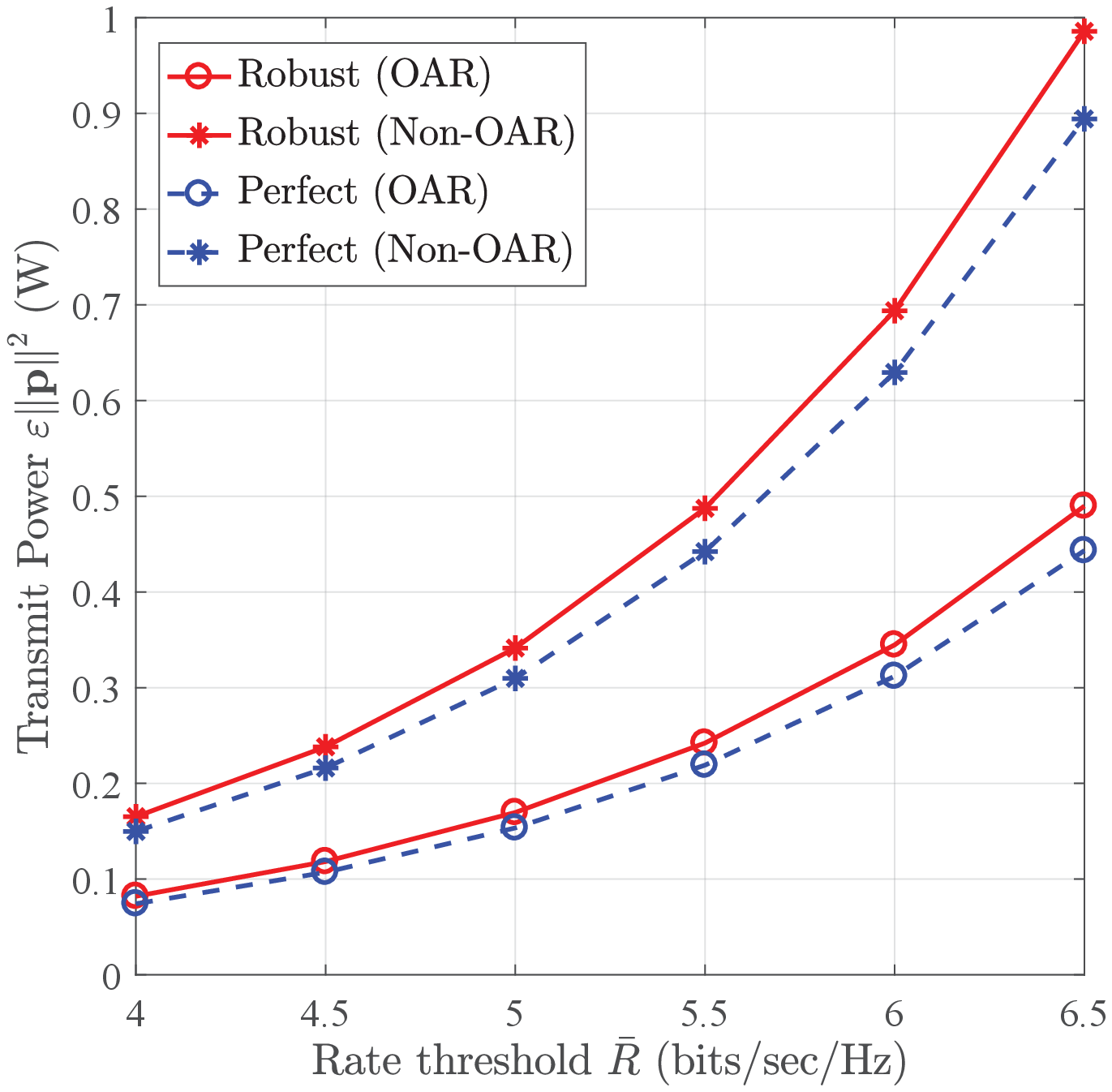}
      \vskip-0.2cm\centering {\footnotesize (a)}
    \end{minipage}\hfill
    \begin{minipage}[b]{0.45\textwidth}
      \centering
\includegraphics[height=7.5cm,width=7.5cm]{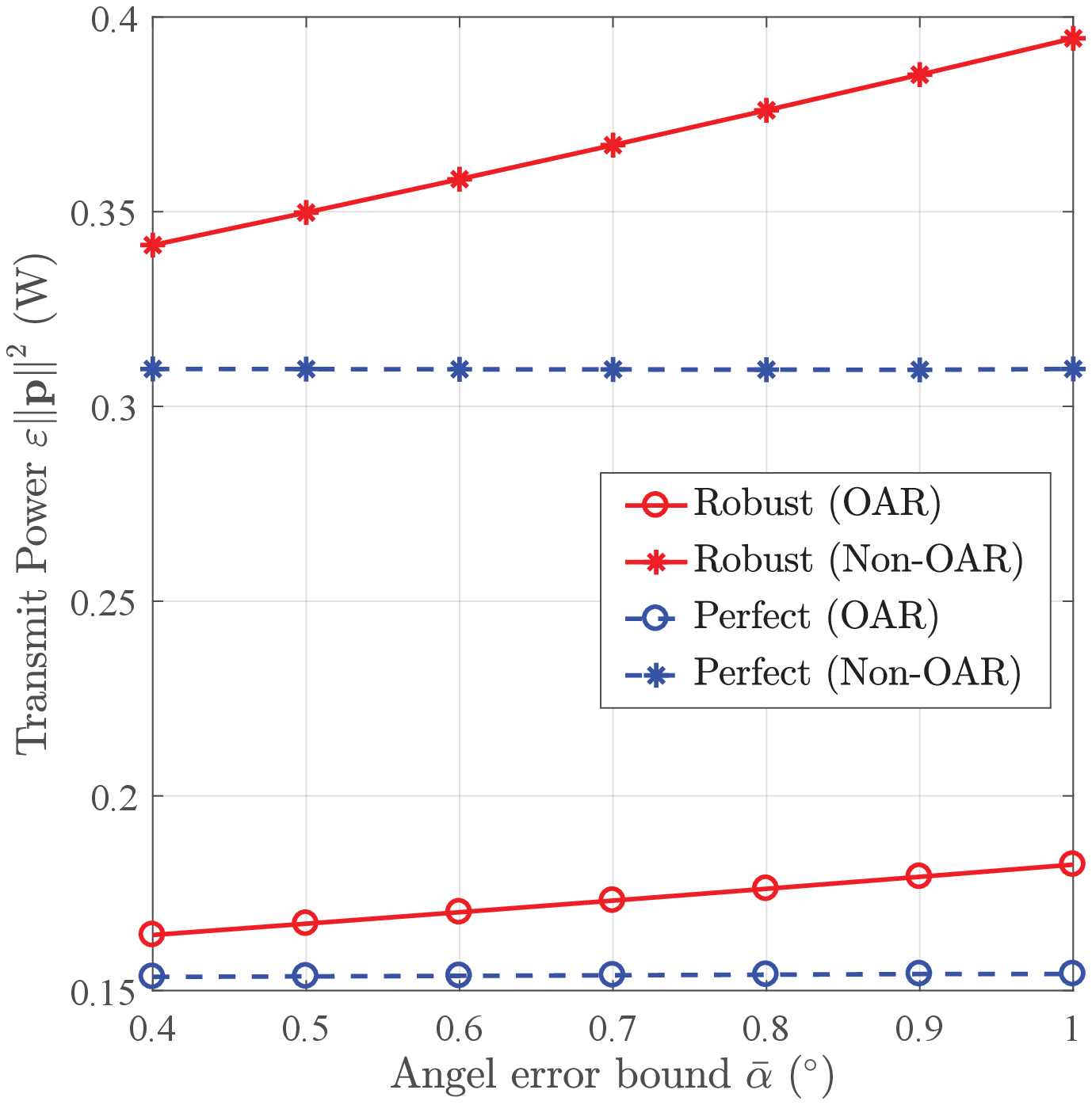}
      \vskip-0.2cm\centering {\footnotesize (b)}
    \end{minipage}\hfill
        \begin{minipage}[b]{0.45\textwidth}
      \centering
\includegraphics[height=7.5cm,width=7.5cm]{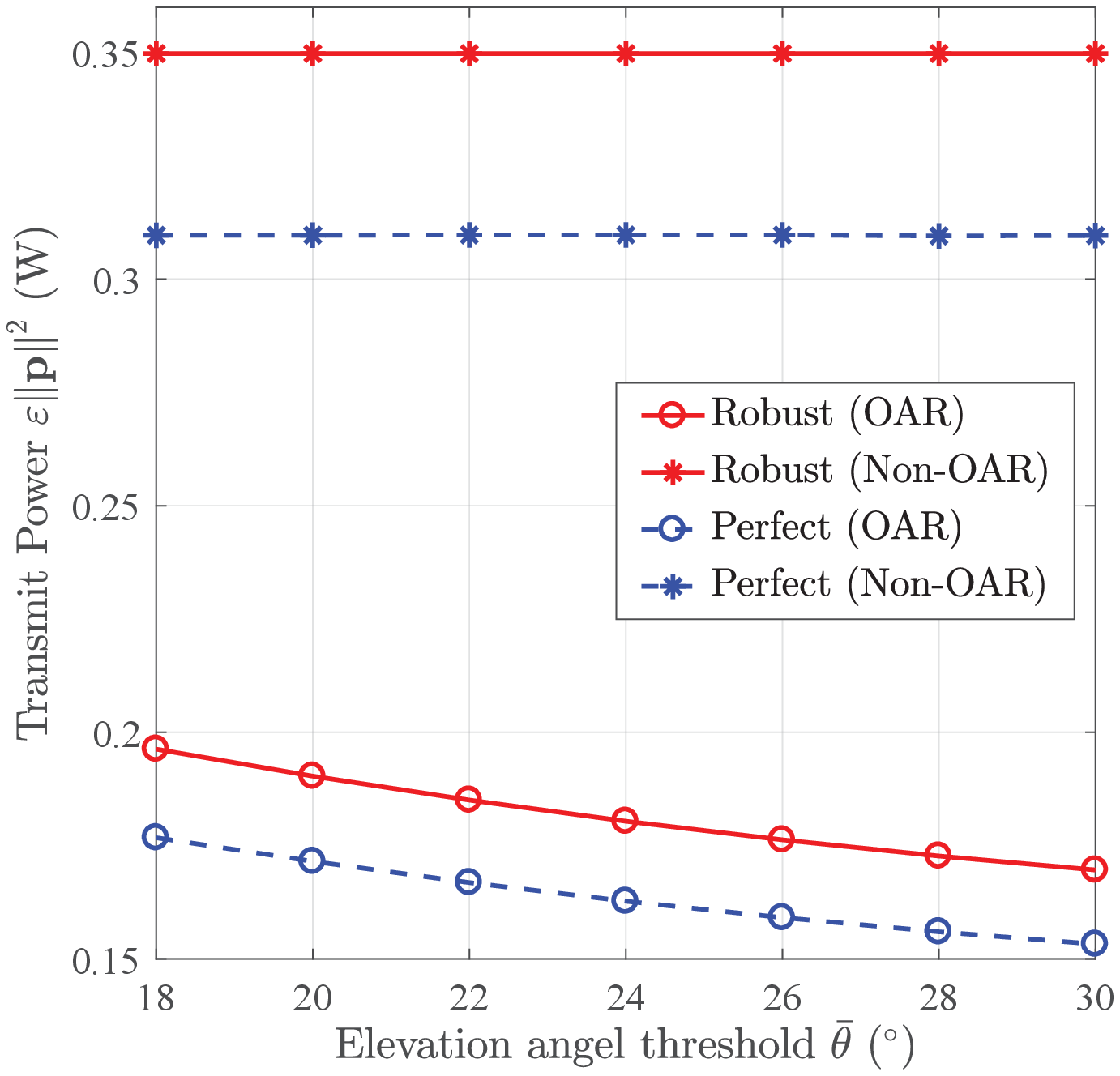}
      \vskip-0.2cm\centering {\footnotesize (c)}
    \end{minipage}\hfill
 \caption{(a)~~Transmit power $\varepsilon{\left\| {\bf{p}} \right\|^2}$ versus minimum rate threshold ${\bar R}$;  (b)~Transmit power $\varepsilon{\left\| {\bf{p}} \right\|^2}$ versus random  angle error bound $\bar \alpha  $; (c)~Transmit power $\varepsilon{\left\| {\bf{p}} \right\|^2}$ versus elevation angle threshold $\bar \theta$.}
  \label{robust_3case_graph} %% label for entire figure
\end{figure}

Fig.~\ref{robust_3case_graph} (a) shows the transmit power $\varepsilon{\left\| {\bf{p}} \right\|^2}$ versus minimum rate threshold ${\bar R}$, in the case of elevation angle threshold $\bar \theta=30^\circ$ and random angle errors $\Delta \alpha  \in \left[ { - {{0.5}^ \circ },{{0.5}^ \circ }} \right)$, $ \Delta \beta  \in \left[ { - {{0.5}^ \circ },{{0.5}^ \circ }} \right)$, $ \Delta \gamma  \in \left[ { - {{0.5}^ \circ },{{0.5}^ \circ }} \right)$. We can observe that  the transmit power of the all aforementioned BO optimization schemes increases  with minimum rate threshold. Moreover, the transmit power of the   robust beamforming design is less than that of the robust design with  non-OAR.
Fig.~\ref{robust_3case_graph} (b) illustrates the transmit power $\varepsilon{\left\| {\bf{p}} \right\|^2}$ versus random angle error bound  $\bar \alpha  $, with minimum rate threshold ${\bar R} = 5 \left( { \rm bits/sec/Hz} \right)$ and elevation angle threshold $\bar \theta=30^\circ$. We observe that as the random angle error bound  $\bar \alpha  $ increases, the transmit power of the robust beamforming design for both OAR and non-OAR increases, where the transmit power of the OAR robust beamforming design is less than the robust design with the non-OAR. However, both perfect cases are constant when
the random angle error increases. This is because  a large random angle error may lead to a bad beamformer design, which leads to increasing  the  transmit power $\varepsilon{\left\| {\bf{p}} \right\|^2}$.
Fig.~\ref{robust_3case_graph} (c) shows the transmit power $\varepsilon{\left\| {\bf{p}} \right\|^2}$ versus elevation angle threshold $\bar \theta$, with minimum rate threshold ${\bar R} = 5 \left( { \rm bits/sec/Hz} \right)$ and random angle errors $\Delta \alpha  \in \left[ { - {{0.5}^ \circ },{{0.5}^ \circ }} \right)$, $ \Delta \beta  \in \left[ { - {{0.5}^ \circ },{{0.5}^ \circ }} \right)$, $ \Delta \gamma  \in \left[ { - {{0.5}^ \circ },{{0.5}^ \circ }} \right)$. We observe that as elevation angle threshold $\bar \theta$ increases, the transmit power of OAR UE decreases for  both robust and perfect scenarios.  This is because  OAR UE can provide more channel gains,  which leads to less transmit power.

%In order to assess the performance of the proposed robust beamforming design for the mobile user, we plot the transmit power with different theta threshold, QoS requirement and radius of user's movement range. Specifically, the circle radius  $r_{\rm{c}}=0.15{\rm{m}}$ and the minimum rate requirement is $\bar R = 5 \left( {{\rm{bits/sec/Hz}}} \right)$ in Fig.~\ref{robust_Vpower} (a). We can observe that the transmit power of all mentioned design schemes decreases as the theta threshold ${\bar \theta }$ increases. In Fig.~\ref{robust_Vpower} (b), the theta threshold  $\bar \theta = 20{^\circ}$ and the minimum rate requirement is $\bar R = 5 \left( {{\rm{bits/sec/Hz}}} \right)$. We can see that the transmit power of proposed design schemes decreases as the circle radius $r_{\rm{c}}$ increases. And in Fig.~\ref{robust_Vpower} (c), the theta threshold  $\bar \theta = 20{^\circ}$ and the circle radius  $r_{\rm{c}}=0.15{\rm{m}}$. We observe that the transmit power of all mentioned design schemes decreases as the minimum rate requirement ${\bar R }$ increases.

\section{Conclusions}

In this paper, we investigate the joint BO schemes
by considering the assistance of an adjustable orientation receiver with  fixed  and random UE orientation.
For the fixed    UE
 orientation,
   the joint BO optimization problem is nonconvex and coupled.  We  develop an alternating optimization algorithm to   optimize   the
  transmit  beamforming and  PD  orientation vectors.
For the  random    UE
 orientation, we further propose a robust  alternating optimization algorithm   based on the S-lemma.
   Finally, the performance of the
joint  BO optimization designs are evaluated through numerical
experiments.
  Our examinations
show that proposed joint BO schemes  enhance the QoS of the UE with less  transmit power   consumption compared to benchmark schemes.

\bibliographystyle{IEEE-unsorted}
\bibliographystyle{IEEEtran}
\bibliography{refs0724}

\end{document}